# The Next Generation of Cherenkov Telescopes. A White Paper for INAF[*].

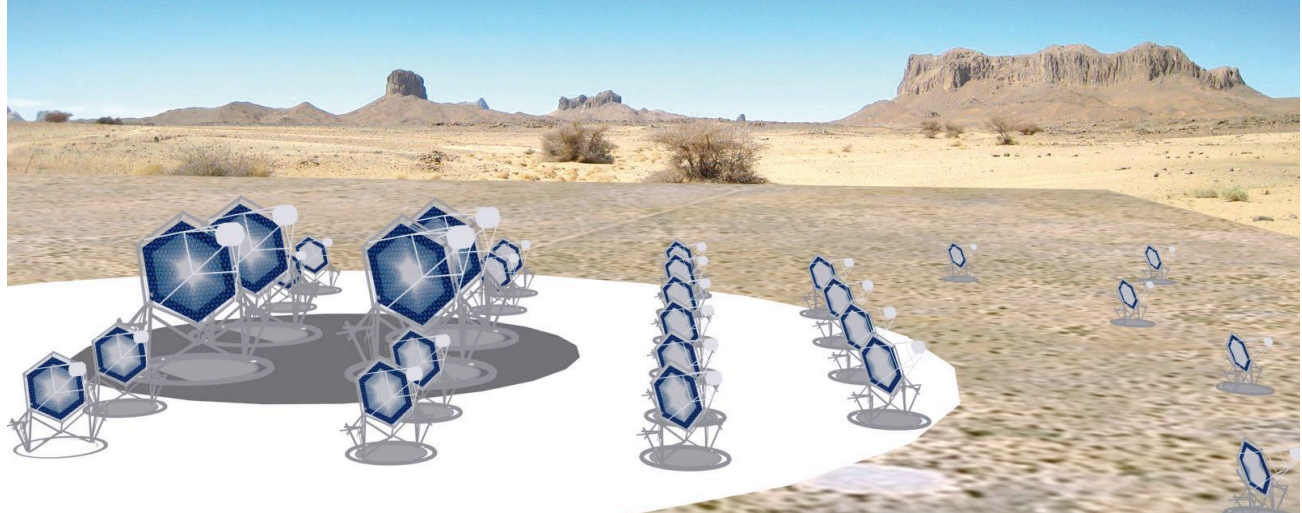

*(Artist view of CTA - courtesy of CTA Collaboration.)*

Prepared by:

L. A. Antonelli[1], P. Blasi[2], G. Bonanno[3], O. Catalano[4], S. Covino[5], A. De Angelis[7],
B. De Lotto[7], M. Ghigo[5], G. Ghisellini[5], G.L. Israel[1], A. La Barbera[4], G. Pareschi[5],
M. Persic[6], M. Roncadelli[8], B. Sacco[4], M. Salvati[2], F. Tavecchio[5], P. Vallania[9]

[1]*INAF-Osservatorio Astronomico di Roma*
[2]*INAF-Osservatorio Astrofisico di Arcetri*
[3]*INAF-Osservatorio Astrofisico di Catania*
[4]*INAF-Istituto di Astrofisica Spaziale e Fisica Cosmica, Palermo*
[5]*INAF-Osservatorio Astronomico di Brera*
[6]*INAF-Osservatorio Astronomico di Trieste*
[7]*Università degli Studi di Udine*, [8]*INFN-Pavia*
[9]*INAF-Istituto di Fisica dello Spazio Interplanetario, Torino*

*\* INAF: Italian National Institute for Astrophysics*

# Contents.

# Executive Summary

In the last decade, thanks to the new generation experiments, Very High Energy (E > 50 GeV) gamma-ray astrophysics has grown into a mature branch of astronomy having increased the number of discovered sources by about a factor of 10. These recent advances of TeV γ-ray astronomy have shown that the 10 GeV – 100 TeV energy band is crucial to investigating the physics prevailing in extreme conditions found in remote cosmic objects as well as to testing fundamental physics. Nevertheless, with the recent launch of two gamma ray dedicated satellites (AGILE and Fermi), the gamma-ray astronomy is now living a sort of *Golden Age* and opening unprecedented opportunities of multiwavelength observations on a very wide energy range. In such an exciting scenario, a new generation of ground-based VHE gamma-ray instruments are needed in order to significantly improve: the sensitivity, the operational bandwidth and the field of view. The international VHE astrophysics community is moving towards such a new generation of Cherenkov experiments both in Europe and USA. Big projects such as the Cherenkov Telescopes Array (CTA) in Europe and the Advanced Gamma-ray Imaging System (AGIS) in USA are now being planned in order to improve the sensitivity of about a factor of 10 and to enlarge the observing band (few tens of GeV up to 100 TeV).

Since 2003, the Italian National Institute for Astrophysics (INAF) has been collecting all Italian research institutes operating in the field of astronomy and astrophysics. INAF scientists and engineers have a long lasting experience in the field of high-energy astrophysics both from space-borne and ground-based experiments. They are involved in space gamma-ray missions such as AGILE and Fermi as well as X-ray missions such as BeppoSAX, INTEGRAL, Newton-XMM, SWIFT. INAF has been also involved in ground-based high-energy experiments (e.g. ARGO-YBJ) and specifically in important Cherenkov telescopes experiment such as MAGIC, GAW. INAF is also realizing and operating several observational facilities observing at all wavelengths.

Relying on this large experience and motivated by the recent challenging results from TeV astronomy, the VHE INAF community asked a group of them to write this White Paper to summarize the status and future of Cherenkov telescopes for γ-ray astronomy and the INAF perspectives in this field. This document wants to review both the scientific topics and potential developments of the field as well as to point out both the interests and the capacities (scientific and technical) of the VHE astrophysics community in INAF. It is aimed at identifying the scientific and technological areas where INAF should focus its efforts and resources so that Italian researchers can achieve (or maintain) a leading position in this field.

Many scientific drivers for Cherenkov telescopes have been identified and discussed in the paper. Some highlights are: *1) Galactic Astrophysics:* Tev γ-ray astronomy can lead to **the identification of the sources of cosmic rays**. In fact, on purely energetic grounds, it is relatively straightforward to single out supernova remnants (SNRs) as the most plausible candidate accelerators. Nevertheless, the only two messengers that would prove this association are gamma rays of unambiguously hadronic origin and neutrinos. Next-generation Cherenkov Telescopes will be the suitable instrumentation for this research allowing us to study in detail the acceleration sites and the propagation of these high-energy particles. *2) Extragalactic Astrophysics:* the exploration at TeV energies of blazars' spectral variability on very small timescales, will allow to establish the **origin and the emission mechanisms of the TeV photons from relativistic jets**. A higher sensitivity in Cherenkov Telescopes will allow the observations of fainter and farthest AGNs and Gamma Ray Bursts (GRB). This will allow us to **measure the extragalactic background light with unprecedented precision**, allowing us to constrain the star formation history of the Universe independently from galaxy counts. Observation of GRBs at these energies will be of fundamental importance to **distinguish between all possible emitting mechanisms** leading to a definitive comprehension of the emitting processes at work in these sources. *3) New Physics:* the next generation of Cherenkov telescopes should have enough sensitivity to observe the γ-ray emission from the self-annihilation of the *neutralino*, which is the currently foremost elementary particle candidate for the cosmological Dark Matter. This will lead to the **direct observation of the *Dark Matter***.

The next generation of Cherenkov Telescopes and the imaging Cherenkov technique will be largely based on a well proven technology. The technological development will move through two different guidelines: 1) optimization in the operation of an array of tens of telescopes having different properties (complex array triggers, observatory operations, robotization, standardization of data format and processing, etc.); 2) enhancements in the performance of the individual telescope (optics, sensors, electronics, signal processing, etc.). INAF is participating in this challenging technological adventure covering different key arguments that reflect the background and expertise present in our institute. In particular, INAF is leader in **mirror technology** (low-cost, low-weight glass mirrors) for large telescopes as well as for the secondary mirror for **large field of view telescopes**. Large field of view may be obtained also with **Fresnel Lenses** following a technology already implemented in the EUSO and GAW projects. Major contributions can also come for an **advanced camera design and electronics** as well as for **trigger systems**. INAF expertise may play a fundamental role also in **data handling, data processing** and for **multiwavelength analysis tools**. A further contribution may also come for the **observatory operations** in particular to the **automatic (robotic) control and operation of the array** as well as for the **Data Center**.

The Cherenkov telescopes arrays and the imaging technique used will be based on a solid and well proven technology so, a factor of 10 in sensitivity can be achieved with an installation cost of the array of about 150 Meuros. This cost represents a small part of the typical budget of a large space observatory or of a next generation big optical observatory. Nevertheless, a relatively low-cost investment can guaranteed an outstanding scientific return to the whole astrophysical community by solving fundamental astrophysical questions and stimulating and driving multiwavelength and multimessenger studies. The INAF participation to these VHE astrophysics projects is strongly recommendable: with a relatively low level of funding a large scientific and technical return will be guaranteed to a larger community. This is also recommended in the Long Term Plan[1] of the institute. The INAF participation to these international projects together with INFN will allow to maintain at a very high level the Italian excellence in this field allowing also a possible involvement of the national industry in the realization of instruments, part of them and/or the infrastructures.

---

[1] http://www.inaf.it/struttura-organizzativa/cs/plt/inaf-long-term-plan

# 1 Introduction.

## 1.1 *Ground-based Gamma-ray Astronomy: Historical Overview.*

High-energy gamma rays are probing the "non-thermal" Universe. They can be produced by all those acceleration processes at work in extreme conditions that can be found in the proximity of black holes or in the very energetic shock waves created in stellar explosions. Otherwise they can be obtained from decays of heavy particles such as the hypothetical dark matter particles or cosmic strings both relics which might be left over from the Big Bang. The flux and energy spectrum of the observed gamma rays bring important information on the emission processes and the physics producing them. Gamma rays at MeV-GeV energies have been typically observed with space-based instruments but at higher energies those instruments are completely unusable. With the advent of the Imaging Atmospheric Cherenkov Telescopes (IACTs) in late 1980's, ground-based observation of TeV gamma-rays came into reality and, since the first source detected at TeV energies in 1989 the number of gamma-ray sources has rapidly grown up to over eighty now as shown in Fig. 1.1.1 (see e.g. [1,2,3] for an extended review).

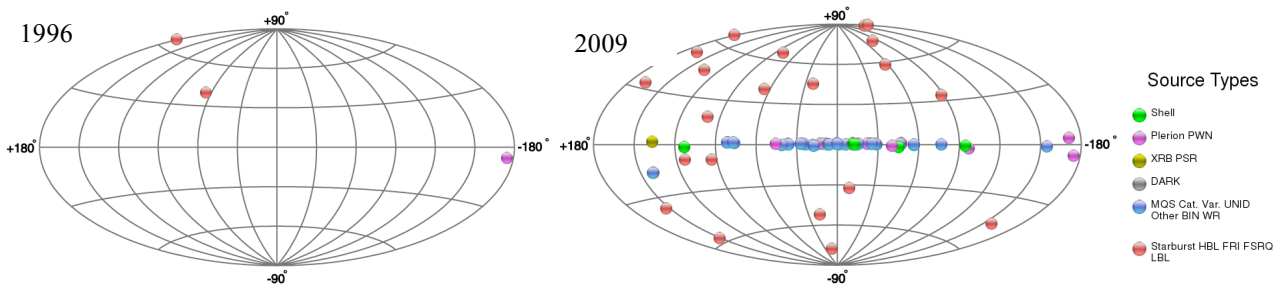

***Fig. 1.1.1:*** *The improvement in the VHE astrophysics from 1996 (left panel) to 2009 (right panel)[2].*

The Imaging Atmospheric Cherenkov Technique for the detection of very high energy (VHE) gamma rays (in the energy range 100 GeV - 10 TeV) was first pioneered by the *Whipple* experiment since 1985 leading to the discovery of TeV gamma-rays from the Crab Nebula in 1989 [4]. This first result was followed by the discovery of the TeV emission from the first extragalactic source (Mrk 421) [5], showing that acceleration processes are taking part in AGNs too. The third source, discovered in 1996, was still an extragalactic object (Mrk 501) [6] which showed a violent flaring activity observed by the European experiment HEGRA [7]. The recent discovery of flux variability on the time scale of few minutes from Mrk 501 and PKS 2155-304, obtained in 2007 by MAGIC [8] and by HESS [9] respectively, has shown that the observed γ-rays are coming from the innermost region of the central part of the active galactic nuclei giving important information on the physical processes at work. The discovery of TeV emission from extragalactic objects was of fundamental importance to constrain the density of the Extragalactic Background Light and the transparency of the Universe to TeV photons [10] (see also par. 3.2). In 2002 HEGRA discovered also the first unidentified TeV γ–ray source [11] showing for the first time that some of the celestial objects discovered at these wavelength emit most of their radiation at VHE energies, or are not detectable in any other waveband.

Since 2003, as the new generation experiments (HESS, MAGIC, CANGAROO and VERITAS) has been started to observe the gamma-ray sky, the number of VHE sources started to rapidly increase. New class of sources was detected at GeV-TeV energies both galactic (e.g. Galactic Center, Pulsar

---

[2] Images created using http://tevcat.uchicago.edu/

Wind Nebulae, Pulsars and Binary Systems [see sect. 2]) and extragalactic (e.g. Blazars, radiogalaxies, star-forming galaxies [see sect. 3]) as well as about a dozen of unknown new TeV sources. The survey of the galactic plane performed by HESS [12] is absolutely remarkable revealing a large population of sources including Pulsar Wind Nebulae and a considerable number of unidentified sources. It showed for the first time that an array of IACTs could be properly used as a real astronomical observatory able to survey a large portion of the sky with a high sensitivity. Among the most outstanding results obtained so far by TeV astronomy there is the recent discovery of pulsed γ-ray emission from Crab Pulsar by MAGIC [13]. This is a very important result providing a unique insight into the structure of pulsar magnetospheres and the main energy transfer processes at work. In March 2007, the HESS project was awarded the Descartes Research Prize of the European Commission for offering "A new glimpse at the highest-energy Universe". Thanks to the two experiment HESS and MAGIC, and to their forthcoming follow-ups HESS 2 and MAGIC 2, VHE astrophysics European community is now firmly leader in this research field.

## 1.2   The Imaging Atmospheric Cherenkov Technique.

The field of ground-based gamma astronomy has been largely driven (with the exception of the remarkable results from MILAGRO) by the exceptional results obtained with the imaging air Cherenkov telescopes (IACTs). As any other optical or radio telescopes, an IACT consists of three basic elements: a mechanical tracking system, which compensate the Earth's rotation, a collecting surface, which gathers the incident electromagnetic radiation and focuses it, and a receiver element, which converts the collected light in a recordable image of the observed field of view (FOV) (see [2] and [14] for a review).

A peculiar feature of Cherenkov telescopes is that they do not detect directly the photon (γ-ray) flux, but instead detect the Cherenkov light produced in the air shower induced by the primary photon. Extensive air showers (EAS) emit in the forward-direction a beam of atmospheric Cherenkov light with a half opening angle of ~ 1°. This beam illuminates almost homogeneously an elliptical region (light pool) on the ground with an area of the order of $10^5$ m$^2$ (depending on the altitude and inclination of the shower axis). An optical telescope pointing to the source and located within the illuminated footprint of the shower can make an image of the air shower against the background light of the night sky, provided the camera is sufficiently fast (~ ns) to integrate the short Cherenkov flash.

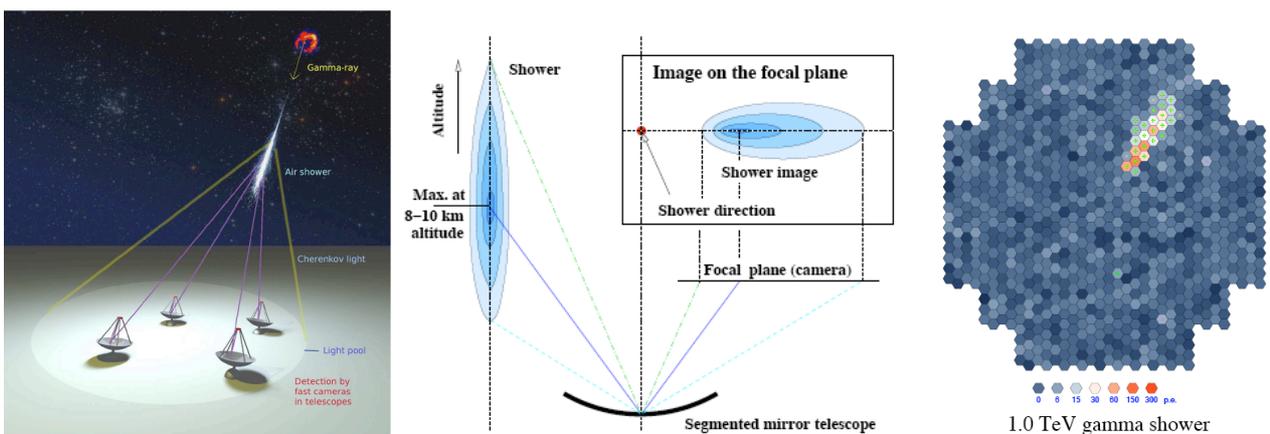

***Fig 1.2.2:*** *Left: Schematic of the Cherenkov light pool, originating from a primary γ-ray and illuminating an array of telescopes. Central: Shower imaged by a telescope. The shower image in the camera has an elliptical shape and the shower direction lays on the extension of its major axis, the image intensity is related to the primary energy. Right: Image of a γ-induced air shower in the camera. (From [2])*

The Cherenkov technique takes advantage of the shower development information in the image of the telescope camera. It is therefore possible to take a sort of "snapshot" of air showers resolved in space (and time). This information can then be used to distinguish the origin of the air shower (hadronic or γ-ray) using the different spatial development of γ- and hadron-induced air showers. The parameterization of such images is called *"Imaging Technique"*, which dramatically improves the γ/hadron separation power and makes IACTs the most successful instrument for cosmic very high energy γ-ray observations. Moreover, the measurement of the Cherenkov light provides a good indicator of the energy absorbed in the atmosphere, which is in fact acting as a calorimeter. Therefore, the total amount of light contained in the image gives the energy of the primary particle. In addition, orientation and shape of the image also provide information on the incoming direction of the primary particle. Two main parameters characterize an IACT: its sensitivity, i.e., the minimum detectable γ-ray flux in a given number of observation hours (usually defined as a 5 σ excess during 50 hours of observation time), and its energy threshold. These parameters are improved considerably when more than one telescope is used in a stereoscopic set-up, with the telescopes separated by tenths up-to hundreds of meters to provide a baseline for triangulating the atmospheric air shower. The stereoscopic technique has become the nominal standard for all current and future installations.

## 1.3 The current status of IACTs.

The current generation of IACT instruments MAGIC and VERITAS (in the northern hemisphere) and HESS and CANGAROO (in the southern hemisphere) (see Fig.1.3.2) are now allowing imaging, photometry and spectroscopy of sources of high-energy radiation with good sensitivity and good angular resolution (see Table 1.3.1).

These experiments are typically working in an energy range spanning between 50-100 GeV to about 100 TeV. The performance of these telescopes is typically characterized by the sensitivity to detect VHE sources with an energy flux down to $10^{-13}$ ergs cm$^{-2}$ s$^{-1}$ in 50 hrs of observation time. This corresponds to a minimum detectable luminosity of $L_{min} \sim 10^{31}$ ergs s$^{-1}$ for a galactic source at a distance of 1 kpc or $L_{min} \sim 10^{41}$ ergs s$^{-1}$ for an extragalactic source at a distance of 100 Mpc. The angular resolution of each reconstructed primary γ-ray is typically better than few arcmin. The relative energy resolution is comparably good and reaches values of ΔE/E ~10-20%. A good picture of the increased sensitivity of these instruments can be obtained looking at the number of new sources discovered from 2004 up-to now compared with the number of sources discovered until 2003 (see Fig. 1.3.1).

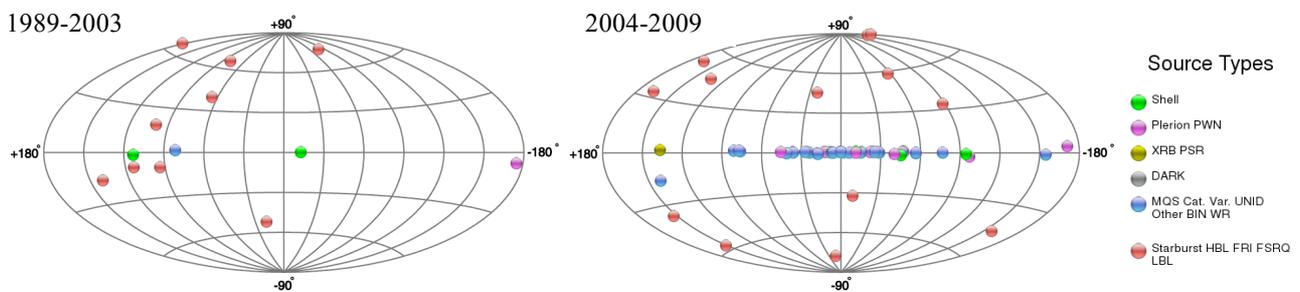

***Fig. 1.3.1:*** *VHE sources discovered until 2003 (left panel) and in the period 2004-2009 (right panel)[3].*

---

[3] image obtained from http://tevcat.uchicago.edu

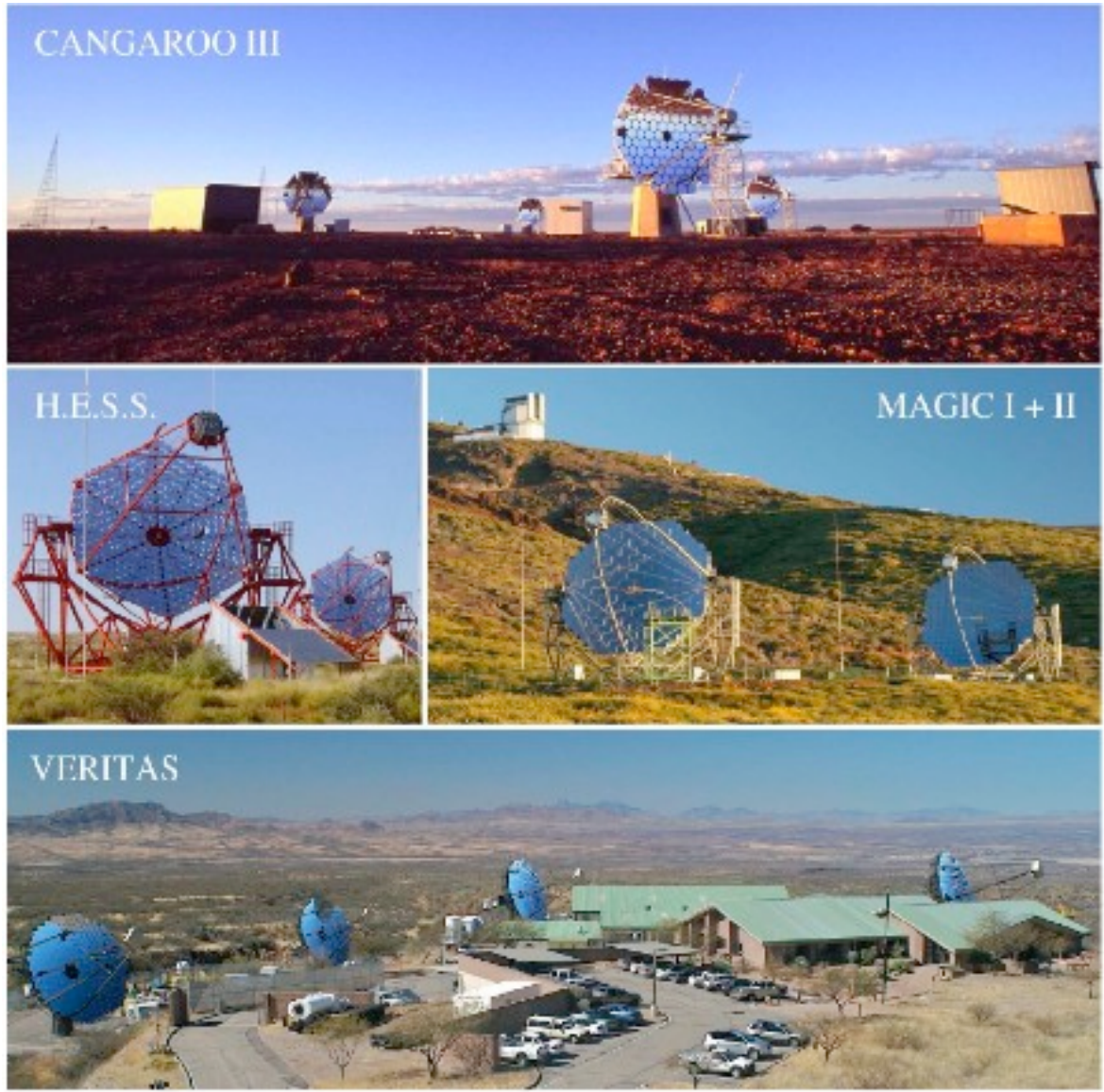

***Fig. 1.3.2:*** *The current generation of IACT instruments: CANGAROO and HESS (in the southern hemisphere), MAGIC and VERITAS (in the northern hemisphere).[4]*

|  | **HESS** | **MAGIC** | **VERITAS** | **CANGAROO** |
|---|---|---|---|---|
| **Location** | Namibia | Canary Islands | Arizona (USA) | Australia |
| **Latitude** | -23° | +29° | +32° | -31° |
| **Current Telescopes** | 4 x 107 m$^2$ | 1 x 240 m$^2$ | 4 x 110 m$^2$ | 4 x 50 m$^2$ |
| **Operational since** | 2004 | 2004 | 2004 | 2007 |
| **Field-of-View** | 5° | 3.5° | 3.5° | 4° |
| **Threshold Energy** | 100 GeV | < 50 GeV | 100 GeV | 400 GeV |
| **Sensitivity** *(5σ in 50h obs. time)* | 0.7 Crab | 1.6 Crab | 1 Crab | 1.5 Crab |
| **Upgrades in progress** | +1 x 600 m$^2$ | +1 x 240 m$^2$ | - | - |

***Table 1.3.1:*** *Characteristics of the current generation of Cherenkov Telescopes.*

---

[4] MAGIC: http://wwwmagic.mppmu.mpg.de/
H.E.S.S.: http://www.mpi-hd.mpg.de/hfm/HESS/
VERITAS: http://veritas.sao.arizona.edu/
CANGAROO: http://icrhp9.icrr.u-tokyo.ac.jp/

The extensions of MAGIC [15] and HESS [16] into phase II since 2009-2010 will lower the energy threshold and improve existing sensitivity. MAGIC II is expected to be in operation by spring 2009. Stereoscopic observations will allow an increase in the sensitivity by at least a factor of 3 as well as further improvements in the energy and direction reconstruction. With the advent of the second telescope, MAGIC will reach a level of 1% Crab Unit in 50 hours of data taking (see Fig. 1.4.1).

The MAGIC experiment has confirmed that a single large telescope can reach a low energy threshold with a good sensitivity. HESS II is following the MAGIC strategy. A large telescope of about 500 tons with a mirror area of about 600 $m^2$ and a camera with a Field of View (FoV) of 3.5° and pixel size of 0.07° is scheduled to be installed in the centre of the current HESS I array by 2010. The HESS II experiment will lower the energy threshold to about 20 GeV.

## 1.4 *The future developments of IACTs.*

The results of the current generation of ground-based gamma-ray instruments such as H.E.S.S., MAGIC, CANGAROO or VERITAS, have shown that the very high energy (VHE) gamma-ray astronomy has grown to a genuine branch of astronomy. Nevertheless, the next decade can be considered the "golden age" of the gamma-ray astronomy with two gamma ray dedicated satellites (AGILE and Fermi) in orbit. Moreover, thanks to many X-ray experiments already in orbit (e.g. Swift, Chandra, Newton-XMM, etc.) and to many other new ground-based optical and infrared instruments, it will be possible to observe the Universe for the first time all over the electromagnetic spectrum almost at the same time. In such a scenario a new generation of ground-based VHE gamma-ray instruments is needed in order to significantly improve the sensitivity, the observed energy band, the field of view, the signal sampling and to reduce the observing time. The VHE astrophysics community is moving towards such a new generation of Cherenkov esperiments both in Europe and USA. Big projects such as the Cherenkov Telescopes Array (CTA)[5] in Europe and the Advanced Gamma-ray Imaging System (AGIS)[6] in USA are now being planned. It is also possible that the efforts currently ongoing in Europe, USA and Japan may unify into a world-wide collaboration to create a gamma-ray observatory capable of addressing the needs of the astronomical community for the next decades.

The European CTA is conceived to allow both detection and in-depth study of large samples of known source types, and to explore a wide range of classes of suspected gamma-ray emitters beyond the sensitivity of current instruments. CTA will be a combination of the well proven technology of Cherenkov telescopes (with some tens deployed over a large area) and of new wide-field gamma detectors. CTA characteristics and performances are described in the following:

• **Sensitivity:** CTA will be about a factor 10 more sensitive than any existing instrument (see Fig. 1.4.1). In its core energy range, from about 100 GeV to several TeV, CTA will have milli-Crab sensitivity in 50h of data taking, a factor 1000 below the strength of the strongest steady sources of very-high-energy gamma rays, and a factor 10000 below the highest flux measured in bursts. This dynamic range will allow the study of weaker and new type sources, reducing the selection bias in the taxonomy of known source types.

• **Energy range:** CTA is aiming to cover, with a single facility, three to four orders of magnitude in energy range (see Fig. 1.4.1). This will enable to distinguish between key hypotheses such as the electronic or hadronic origin of highest energy gamma rays. Combined with the GLAST satellite gamma-ray observatory, the two instruments will provide an unprecedented seamless coverage of more than seven orders of magnitude in energy.

---

[5] http://www.cta-observatory.org/
[6] http://www.agis-observatory.org/

• **Angular resolution:** Current instruments are able to resolve extended sources, but they cannot probe the fine structures visible in other wavebands. Selecting a subset of gamma ray induced cascades detected simultaneously by many telescopes, CTA will reach angular resolutions in the arc-minute range, a factor 5 lower than current instruments (Fig. 1.4.2 right panel).

• **Temporal resolution:** With its large detection area, CTA will resolve flaring and time-variable emission on sub-minute time scales which are currently not accessible.

• **Flexibility:** Consisting of a large number of individual telescopes, CTA can be operated in a wide range of configurations, allowing on the one hand the in-depth study of individual objects with unprecedented sensitivity, on the other hand the simultaneous monitoring of tens of objects (relevant for flaring sources) and any combination in between (Fig. 1.4.2 left panel).

• **Survey capability:** Groups of telescopes can point at adjacent fields in the sky, with their fields of view overlapping, providing an increase of sky area surveyed per time unit by an order of magnitude, and for the first time enabling a full-sky survey at high sensitivity.

• **Number of sources:** Extrapolating from the intensity distribution of known sources, CTA is expected to enlarge the catalogue of objects detected from currently about 80 objects to about 1000 objects.

• **Global coverage and integration:** CTA aims to provide global coverage of the sky from multiple observatory sites, using transparent access and identical tools to extract and analyse data.

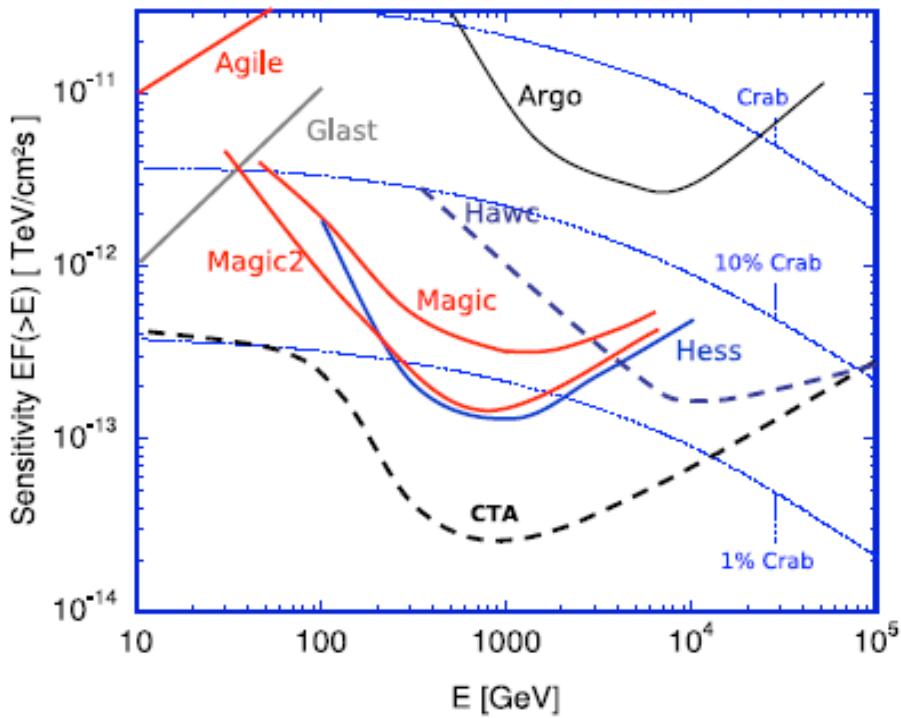

*Fig. 1.4.1: Sensitivities of some present and future HE gamma detector, measured as the minimum intensity source detectable at 5 sigma. The performance for EAS and satellite detector is based on one year of data taking; for Cherenkov telescopes it is based on 50 hours of data (from [3]).*

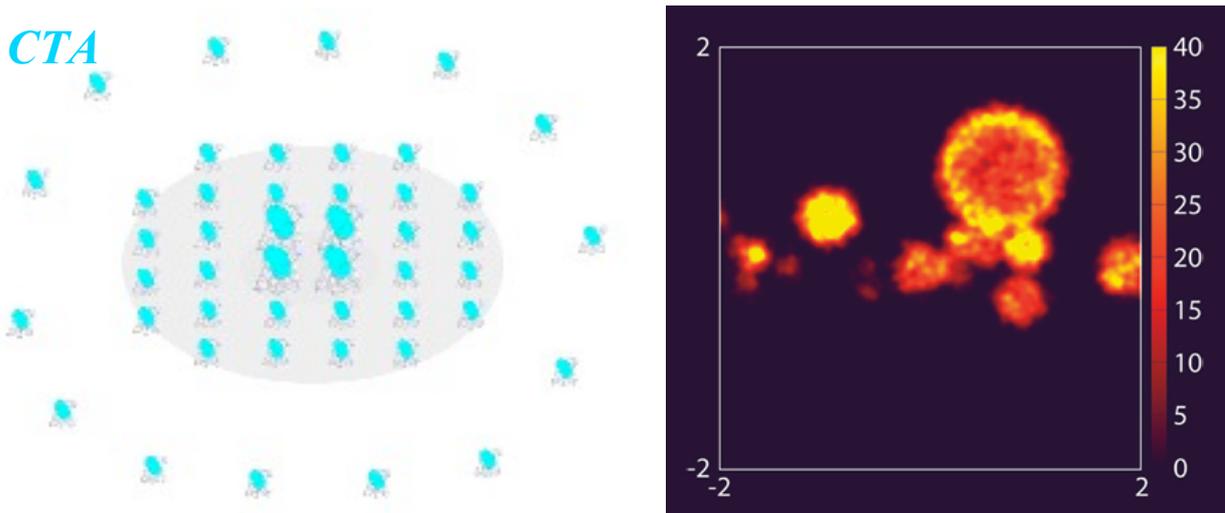

*Fig. 1.4.2: (Left panel) A Cherenkov Telescopes Array possible design scenario. (Right panel) Simulation of the sensitivity and the spatial resolution of CTA in a small (4°x4°) region of the Galactic Plane containing Supernovae Remnants. (courtesy of CTA Collaboration).*

CTA, will be, for the first time in this field, operated as a true observatory, open to the entire astrophysics (and particle physics) community, and providing support for easy access and analysis of data. Data will be made publicly available and will be accessible through Virtual Observatory tools. The large amount of data obtained and open to public access will also allow data mining in addition to targeted observation proposals favouring multiwavelength studies. The CTA project aims to emerge as a cornerstone in a networked multi-wavelength, multi-messenger exploration of the Non-thermal Universe. For this reason also issues such as data reduction and dissemination have to be conceived and developed within the CTA project.

CTA has been included in the 2008 roadmap of the European Strategy Forum on Research Infrastructures (ESFRI)[7]. It is one of the seven most important projects of the European strategy for astroparticle physics published by ASPERA[8], and highly ranked in the "strategic plan for European astronomy" of ASTRONET.

On the US side, the AGIS telescopes array will have similar features as CTA. AGIS was recommended in the *"US Particle Physics: Scientific Opportunities"* report (May 2008). At the end of 2008, following that recommendation (and after having received a specific invitation by the Division of Astrophysics of the American Physical Society), the AGIS collaboration successfully prepared a white paper on the status and future of ground based TeV gamma-ray astronomy [17]. The Particle Astrophysics Scientific Assessment Group (PASAG) has then requested (Feb 2009) to explore in further detail the opportunities and scientific challenges available at the Cosmic Frontier (P5 report), while the AGIS concept has been submitted as a project for the 2010 Decadal Survey of the US National Academy for Physics and Astronomy. It should be noted that INAF it is represented (unique European Institution) in the AGIS project, being the Brera Astronomical Observatory member of the collaboration.

---

[7] ftp://ftp.cordis.europa.eu/pub/esfri/docs/esfri_roadmap_update_2008.pdf
[8] http://www.aspera-eu.org/images/stories/roadmap/aspera_roadmap.pdf

## 1.5 The INAF perspective for the new generation of IACTs.

Since 2003 the Italian National Institute for Astrophysics (INAF) is collecting all the Italian research institutes operating in the field of astronomy and astrophysics (formerly Astronomical Observatories and ex-CNR institutes). INAF has inherited the long lasting experience in the field of the high energy astrophysics of the former institutes. INAF is currently participating to projects in the VHE gamma astrophysics including Cherenkov experiments (e.g. GAW and MAGIC)[18, 19] providing both technological and scientific support. INAF scientists are also participating to the Design Study of the CTA Project [19]. The INAF contribution to CTA is reflecting its background and know-how in the field of VHE astrophysics, VHE technologies, space- and ground-based observatories management and operations, multiwavelength observations and data analysis. In particular, INAF scientists and associated are mainly participating to CTA in the following topics:

*Astrophysics and Astroparticle Physics.* INAF scientists are actively involved in the theoretical investigations in the field of High-Energy Astrophysics. In particular, different groups have focused on particular topics of research of interest for the CTA Design Study (DS). The group at the OA Brera (Milan) are actively working in the development of theoretical models of emission processes in relativistic jets, in the organization of high energy and multifrequency observations of GRB, AGN and Blazars and in modelling their Spectral Energy Distributions. Groups at OA Roma, OA Brera, IASF Roma, IASF Palermo and IASF Milano are active in theoretical modelling and in organizing high energy and multifrequency observations of Galactic Compact Objects and Gamma Ray Burst and in modeling their both spectral and temporal properties. A group at the Arcetri Observatory focuses its research in modeling particle acceleration in SNRs, plerions, relativistic shocks for applications to PSR systems, cosmic-ray origin and propagation models, AGNs, and galaxy clusters. OA Trieste is also involved in studying Dark Matter and fundamental physics.

*Optimisation of the array layout.* The IFSI Torino group pioneered the Cherenkov observations by detecting in 1983 the Cherenkov light reflected from mountain snow at high altitude [20]. Further results were obtained with the Gran Sasso EAS-TOP detector [21] and the ULTRA experiment [22]. The group is currently involved in the ARGO-YBJ experiment, an extensive air shower array dedicated to the study of the GeV-TeV astronomy [23]. In particular they are expert in simulations and studies to characterize and to optimize the instruments performances in the TeV energy band with particular attention to the large field of view telescopes.

*Telescope Optics and Mirrors.* The OA Brera group has a long and successful experience in the development of X-ray astronomical optics realized by replication (SAX, XMM and Swift projects) as well as an experience recently acquired in the development activity of large-size segmented mirrors in thin slumped glass for future large size ground based optical telescopes. The OA Padova group has developed a large experience in the design, calibration and alignment of active optics in ground-based telescopes. Aluminum mirrors of MAGIC have been also realized for the MAGIC telescope at the OA Padova in collaboration with INFN-LNL (Legnaro). Both OA Brera group and OA Padova group have participated to the development and production of the glass mirrors MAGIC 2 telescope [24] and they are both involved in the development of new production techniques for the realization of the CTA mirrors [25]. It particular, in collaboration with industrial partners (Media Lario), they are already involved in the development of reflecting panels based on the new approach of the cold slumping of thin glass sheets from a master, a technique suitable for the mass and low cost production of the CTA mirrors. IASF-Palermo has acquired a relevant experience in ray-tracing from optic systems as well as a deep experience in simulation of Cherenkov light produced by atmospheric Air Shower with particular regard to the wide field. These activities have been carried on in the context of the GAW (Gamma Air Watch) experiment.

***Photon detectors and focal plane.*** A group at IASF-Palermo is active in the design of sensors and relative electronics for high energy detectors operating in both space-borne and ground-based experiments. This group contributed both to the Beppo-SAX satellite and to the EUSO design as well as the GAW experiment. GAW is a Cherenkov experiment based on the use of Fresnel lenses instead of mirrors and on the electronics Single Photon Counting method instead of the more usual signal integration mode. A group in OA Catania has a large experience in the use and qualification of photomultipliers in the UV band. Both groups are interested in participating to CTA with the study of Photon Detectors (PDET). IASF-Palermo is also supporting and realizing a dedicated Breadboard (front-end electronics) for test setup and characterization of photo sensors, and collaborating to the activity for the construction of a partially filled camera prototype for CTA.

***Observatory operation.*** INAF has considerable expertise in running astronomical telescope in Italy, Canary Islands (TNG), Chile (REM) and Arizona (LBT). Several INAF groups have been and are still involved in operating telescopes and instruments, in setting up and running scientific Announcements of Opportunity, in the management of observing proposals and in data distribution from both ground- and space- based observatories. OA Roma group is participating in defining the requirements to operate CTA as an Observatory open to the scientific community. OA Roma and OA Brera are also participating to activities towards an automatic (robotic) control and operation of CTA telescopes and software development for a fail-safe operation and control of the instruments.

***Site selection.*** OA Bologna and IASF-Palermo are participating in the definition of the requirements of the CTA site and infrastructures. OA Bologna is involved in the activity of data calibration in order to correlate ground-based data with satellite data.

***Data handling, data processing, data access.*** INAF has a long lasting experience in astronomical data handling. INAF groups have been strongly involved in data analysis and processing from instruments and experiments at all wavelengths. OA Roma, IASF-Bologna, and OA Brera are contributing to the CTA system design in defining data format, design and development of quick look analysis, pipeline processing and data storage systems as well as to the science analysis tasks.

INAF can give also an important contribution to the AGIS program. In such a context, collaboration (between Brera and UCLA) is already ongoing for the development of the primary segmented mirrors of the AGIS telescope arrays. AGIS foresees also the use of wide field two-mirrors telescopes [26] and INAF is also engaged in the development of the large segmented secondary mirrors via hot slumping [27]. IASF-Palermo will also collaborate with AGIS in the field of new generation photon-counting detectors.

**References:**

[1] Trevor C. W., 2008, arXiv:0811.119v1
[2] Völk H.J. & Bernlöhr K., 2008, arXiv:0812.4198v1
[3] De Angelis A., Mansutti, O. & Persic M., 2008, NCim., 31, 187 (preprint: arXiv: 0712.0315v6)
[4] Weeks et al., 1989, ApJ, 342, 379
[5] Punch, M. et al., 1992, Nature, 358, 477
[6] Quinn, J. et al., 1996, ApJ 456, L83
[7] Aharonian, F. et al., 1997, A&A 327, L5
[8] Albert, J. et al., 2007, ApJ, 669, 862
[9] Aharonian, F. et al., 2007, ApJ 664, L71
[10] Zweerink, J.A. et al., 1997, ApJ 490 L141
[11] Aharonian, F. et al., 2002, A&A, 393, L37
[12] Aharonian, F. et al., 2006, ApJ 636, 777
[13] Aliu, E., et al., 2008, Science, 322, 1221
[14] Lidvansky, A. S., 2006, Radiation Physics and Chemistry, 75, 891
[15] Goebel, F., 2008, Proc. 30th Int. Cosm. Ray Conf., UNAM, 3, 1485
[16] Vincent, P., 2006, Proc. 28th Int. Cosm. Ray Conf., Pune, India, 5, 163
[17] Buckley, J., et al., 2008, arXiv0810.0444v1
[18] Cusumano, G. et al., 2008, Proc. 30th Int. Cosm. Ray Conf., UNAM, arXiv0707.4541
[19] Antonelli, L.A., 2008, Mem. SAIt, Proceedings of the "LII congresso della Societa' Astronomica Italiana", Teramo, May 4-8 2008, arXiv0806.4363
[20] C. Castagnoli, G. Navarra and C. Morello, 1983, Il Nuovo Cimento C, Volume 6, Issue 2, p 202.
[21] M. Aglietta et al., 1992, Il Nuovo Cimento A, Volume 105, Issue 12, p. 1806
[22] G. Agnetta et al., 2007, Nuclear Instruments and Methods A, Volume 570, p.22
[23] G. Aielli et al., 2008, Astroparticle Physics Volume 30, p. 85
[24] Pareschi et al., 2008, Proc. SPIE, 7018, 70180W
[25] Vernani, D., et al. 2008, Proc. SPIE 7018, 70180V
[26] Vassiliev, V.V., et al., 2007, Astropart. Phys, 28, 10
[27] Canestrari, R. , et al., 2008, Proc. SPIE, 7018, 70180D

# 2 Galactic Astrophysics.

## 2.1 *Supernova Remnants and the origin of Cosmic Rays.*

Almost one hundred years after the discovery of cosmic rays (CRs) by V. Hess in 1912 we are now at a turning point in the search for the sources of this energetic radiation. On purely energetic grounds it is relatively straightforward to single out supernova remnants (SNRs) as the most plausible candidate accelerators. Nevertheless, the only two processes that would prove this association (i.e. gamma rays of unambiguously hadronic origin and neutrinos), have only recently become potentially detectable. While on energetic grounds the association between CRs and SNR is easy to postulate, one of the most problematic aspect of CR acceleration in SNRs has been that of reaching energies which could be high enough, in that the typical turbulence present in the interstellar medium (ISM) would allow shock acceleration to energize CRs only up to few GeV. On the other hand it has been understood that CRs can in fact self-generate the levels of turbulence needed for acceleration up to $\sim 10^{15}$ eV, through the excitation of an instability ahead of the shock front [1]. The necessary magnetic field is ~100-1000 times larger than that in the ISM. One of the most important observational discoveries of the last few years has come from X-ray observations: Chandra and XMM have shown that the non-thermal X-ray emission from SNRs is spatially very narrow, confined in filaments with a size of typically ~0.01 pc (see [2] for a review). This morphological information leads to the important but yet tentative conclusion that we are observing the electrons in their act of losing energy because of synchrotron losses behind the surface of the shock [3]. This interpretation would lead to conclude that indeed the magnetic field in the acceleration region is of order a few hundred μGauss, the same level required for acceleration up to the knee [4,5]. The interpretation of the X-ray rims is however currently subject of much debate and a definite conclusion should come from a combined understanding of the spectra, morphology and time variability of the X-ray signal.

Another invaluable insight has been provided by gamma ray observations carried out by Cherenkov imaging telescopes. In particular, the HESS detection of gamma rays up to energies in excess of 10 TeV from a few remnants where non-thermal activity at other frequencies are also observed, has strengthened the case in favor of SNRs as plausible CR accelerators. An especially interesting case is that of RXJ1713-3946 [6] where the morphology of the emission is also resolved (see Fig. 2.1.1, left panel). The gamma ray emission from this remnant extends to almost 100 TeV, with a cutoff at ~10 TeV (see Fig. 2.1.1, right panel). Similar gamma ray emission has been observed from RXJ0852.0-4622 [7].

Despite this progress, the case for SNRs as sources of CRs is still circumstantial, though at a level that certainly calls for further investigation. The role of gamma ray astronomy in the TeV region is clearly key to build a case for or against the SNR paradigm. Below we discuss several avenues through which the next generation Cherenkov imaging telescopes, and especially CTA, may lead to progress in the field.

Gamma rays are expected from SNRs, and more in general from CR accelerators, mainly because of two physical mechanisms: production and decay of neutral pions in inelastic collisions of CRs with the ambient gas, and inverse Compton (IC) emission from accelerated electrons against the photons of the cosmic microwave background and the infrared background. The left panel of Fig. 2.1.2 shows the spatially integrated spectral energy distribution of RX J1713.7-3946 in the hadronic scenario: the solid lines refers to the synchrotron emission, the dotted line is the thermal emission assuming that electrons and protons share the same temperature. Compton scattering with CMB (dashed line) and with Opt+IR background (dot-dashed line) are also shown. The contribution from pion decay is shown as a thick solid line and corresponds to $E_{max} = 1.26 \; 10^5$ GeV. HESS data taken from 2003 to 2005 are plotted together with Suzaku data in the X-ray band. Also EGRET upper limit and GLAST sensitivity are shown.

The right panel of Fig. 2.1.2 shows the spatially integrated spectral energy distribution of RX J1713.7-3946 in the leptonic scenario. The following components are plotted: synchrotron (thin solid line) and thermal (thick dotted line) electron emission, ICS component for CMB (dashed line), optical (dotted line) and IR (dot-dashed line) photons, and the sum of the three (thick solid line). The Opt+IR components must be assumed to have energy density 24 times the mean ISM value in order to fit HESS data.

The Chandra detection of non-thermal X-ray emission from SNRs is solid proof that electrons are accelerated in these sources up to energies of at least several TeV. The X-ray emission is produced by electrons close to their maximum energy through synchrotron emission in the local magnetic field. The shape of the cutoff spectrum of accelerated electrons depends on whether the maximum energy is determined by energy losses or by finite age of the accelerator, so does the spectrum of the X-ray emission and of the IC gamma ray emission. An accurate measurement of the X-ray emission, already at reach with Suzaku (see [8] and X-ray data points in Fig. 2.1.2), and of the gamma ray emission at the highest energies can therefore provide us with a clue to the physical processes that originate the gamma ray emission. The gamma ray spectrum is in fact again somewhat different if of hadronic origin: in the context of the non-linear theory of particle acceleration at the shock [9], the spectrum of accelerated CRs has some degree of concavity which reflects in the spectrum of the secondary gamma radiation. A discrimination between a leptonic and a hadronic origin in the 1-100 TeV regime requires a statistical + systematic uncertainty in the energy determination of the photons better than 30%. A future Cherenkov gamma ray telescope would be required to have a wide response in energy so to cover energies below TeV as well. This would lead to the potential discovery of the concavity mentioned above, one of the most distinctive features signaling for efficient CR acceleration in SNRs.

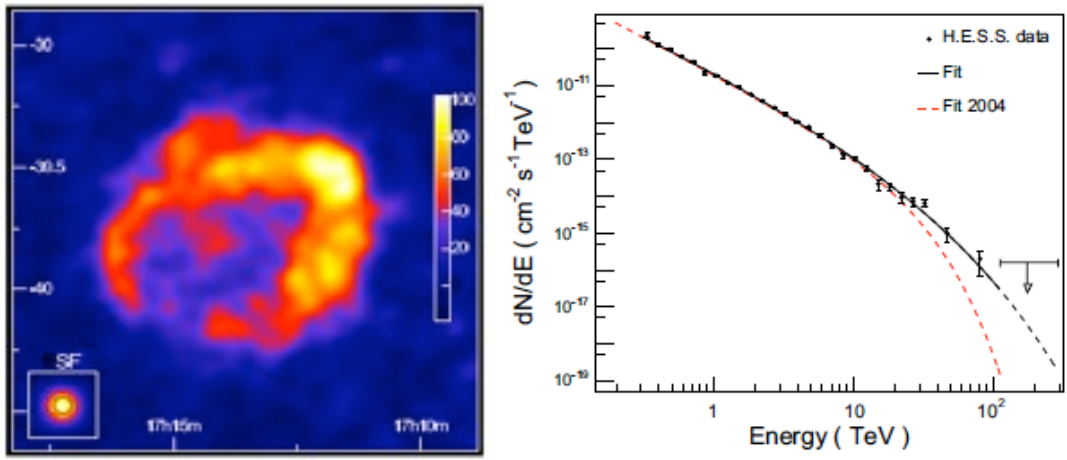

*Fig. 2.1.1: **Left**: HESS image of RXJ1713-3946; **Right**: Gamma ray spectrum as measured by HESS [6]*

An important point to realize is that despite the fact that we expect SNRs might accelerate CR protons up to about 1 PeV, the gamma ray emission extends only up to 100 TeV. But even in this case, only very few SNRs will show a gamma ray spectrum with a cutoff at such high energies: the highest maximum energy is reached at the very beginning of the Sedov-Taylor phase and this stage lasts a relatively short time, after which the maximum energy decreases with time and so does the cutoff energy in the gamma ray spectrum. It follows that it is very unlikely to catch a SNR in the act of accelerating CRs up to the knee. However, a telescope with high sensitivity, like CTA, may detect the gamma ray emission from a large number of SNRs thereby providing different snapshots in time of the acceleration history of SNRs. This would be a precious piece of information to make a case for or against these sources as accelerators of CRs.

As mentioned above, it would be desirable for a next generation Cherenkov telescope to cover as much as possible of the energy region 10-100 GeV. This would have three major positive implications: 1) detection of a possible concavity in the spectrum; 2) connection with the energy region covered by Fermi/LAT; 3) unambiguous discrimination between leptonic and hadronic models. The third point is especially important: as visible in Fig. 2.1.2, even if both leptonic and hadronic models may potentially explain data in the HESS region, the predictions diverge quite remarkably at lower energies: there the IC spectrum is much flatter and close to the sensitivity of Fermi/LAT while hadronic models predict a larger gamma ray flux.

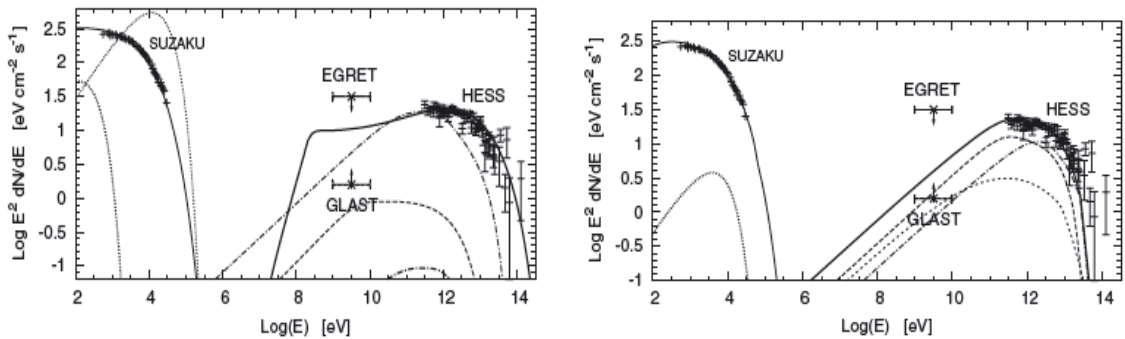

*Fig. 2.1.2: Multifrequency spectrum from RXJ1713-3946 [10]. **Left:** X-ray emission and gamma ray emission in the hadronic scenario for RXJ1713-3946. **Right:** Leptonic scenario for RXJ1713-3946.*

One of the most interesting and harder problems in cosmic ray physics is to understand the mechanisms responsible for the escape of CRs from their accelerators. CRs accelerated at a SNR are dominantly confined within the remnant with the exception of particles very close to the maximum achievable at any given time. During the Sedov-Taylor phase these few particles can escape the accelerator. The rest of the particles can only escape after the breakdown of the shock wave, when however the particles will have lost an appreciable fraction of their energy through adiabatic losses. The flux of CRs that we observe at Earth is the convolution in time of these two components, and is therefore not related in any immediate way to the spectrum of CRs that we can infer from direct gamma ray observations of a SNR. This phenomenon is utterly important if we wish to establish a connection between SNRs and CRs and TeV gamma ray observations may provide us with a crucial piece of information: particles escaping towards upstream of the shock wave may occasionally impact on a nearby molecular cloud producing gamma rays because of inelastic hadronic collisions [11,12]. The spectrum of gamma rays arising from this process is peculiarly flat and might be the only direct way to catch CRs in their act of escaping the parent accelerator.

The next generation of Cherenkov gamma ray telescopes will operate at a very special time, in that for the first time they will be in the right conditions to identify the sources of cosmic rays in the Galaxy, because of their sensitivity, of their expected angular resolution and the energy range they are expected to cover. But also because they may be expected to be operating at qualitatively the same time that gigantic neutrino telescopes should be at work, providing the missing neutrino signal from the birth cry of CRs.

## 2.2 Pulsar Wind Nebulae.

PWNe are astrophysical sources powered by the rotation of a central pulsar that is responsible for the production of a magnetized relativistic wind, mainly made of electron-positron pairs. The particles in the wind are isotropized at a highly relativistic termination shock, where particle acceleration takes place and particles reach extremely large Lorentz factors, in excess of $10^8$. The propagation and losses of electrons in the surrounding medium lead to the diffuse emission extending from the radio to the gamma-ray band that is observed as a PWN.

Several PWNe have been observed (and some of them have even been discovered by HESS): these objects are the main class of galactic TeV sources by number of detections. The Crab is the obvious stereotypical case of a PWN, but HESS has also detected, among the most famous cases, PSR B1823-13 and Vela X [13], and resolved the first jetted emission from a PWN in the TeV range in the PWN associated with PSR1509-58 [14].

The bulk of the emission from PWNe is usually interpreted as the result of synchrotron and IC emission from ultra-relativistic electrons, the former being usually cut in the MeV energy range. The leptonic interpretation of the TeV emission is however far from established in general. In some PWNe one can "see" the effects of aging of the electron population (spectral steepening as a function of distance from the acceleration region) and the consequent morphology, while in other cases the leptonic scenario is not tested. In all cases comparison between high-quality observations and detailed gamma-ray modelling can be used to discover or at least impose limits on the presence of hadrons in the relativistic wind. Far from being just a curiosity, hadrons appear to be needed in some models of particle acceleration at the termination shock [15], and might be the charge carriers providing current closure for the pulsar circuit. The implications of the presence of hadrons in PWNe for gamma ray and neutrino astronomy were discussed by [16].

HESS observations have also provided unexpected news on the morphology of some PWNe, such as HESS J1804-216 [17], HESS J1825-137 [18] and HESS J1718-385 [19], for which the pulsar appears to be located far from the centroid of the nebulae. This has led to think that these are cases in which the ambient medium is very inhomogeneous and the interaction of the reverse shock with the nebulae in some directions changes the appearance of the PWN.

Clearly the main improvement in our understanding of these complex objects, as already happened in X-rays, will come from a better angular resolution with future Cherenkov telescopes and more specifically with CTA, so to provide imaging of the emission in fine details. In this sense PWNe are ideal laboratories to study particle acceleration at relativistic shocks, which is crucial not only for PWNe but also for a variety of other astrophysical objects.

## 2.3 Pulsar.

The periodic emission of pulsars is generally attributed to electro-optical cascades that should originate in regions with non-screened, parallel electric field (the so called "gaps"), and then develop through the magnetosphere (see [20] for a review). This basic scheme has however a vast number of ramifications and uncertainties, connected with the details of the acceleration and the emission processes which actually govern the development of the cascade.

A key ingredient is the geometric location of the gaps: we have the extreme cases of "polar" gaps, located just above the neutron star surface at the feet of the open field lines; and the "outer" gaps, located outside the zero-charge-density surface and extending up to the light cylinder. Over the years, various intermediate cases have been proposed and investigated, so that the original clear-cut dichotomy is now somewhat blurred.

Both observation and theory indicate that the pulsed spectrum terminates somewhere between several GeV and several tens of GeV, i.e., just between the typical bandpass of gamma-ray satellites and that of atmospheric Cherenkov telescopes.

Theoretically, the exact value of the cut-off energy and the precise shape of the cut-off depend -of course- on the maximum energy where acceleration and radiative losses reach equilibrium, and also on the propagation effects across the magnetosphere. In a large region of the parameter space the propagation effects are the dominant ones: the primary gamma rays are absorbed by magnetic pair production in the intense pulsar field, and by photon pair production in the circum-pulsar radiation field. Qualitatively speaking, the first mechanism should dominate in the polar gap scenario, giving rise to a super-exponential cut-off; the second mechanism could be dominant only in the outer gap scenario, and would give rise to a milder, (sub) exponential rollover.

Observationally, the very well studied object, the Crab pulsar, was until very recently without a measured cut-off: it was detected with no sign of downturn at GeV energies by EGRET [21] (and now by Fermi), and was undetected with tight upper limits at several tens of GeV. The cut-off has been finally measured by MAGIC, in an observing mode with a threshold of only 25 GeV, unprecedented for IACTs [22]. The relatively high cut-off energy and the spectral shape are in favor of an outer gap scenario in this object.

Other GeV-bright pulsars, e.g., Vela and Geminga [21], exhibit a downturn at GeV energies, however the remaining satellite band-pass is too narrow to allow the measure of the cut-off shape. Moreover, Fermi is now finding new GeV pulsars. A CTA-like instrument, with large effective area at low energies, would find here a rich and fruitful field to investigate.

## *2.4 Binary Systems.*

By the end of the EGRET era no X-ray binary system was recognized to be a gamma ray emitter in and beyond the MeV band. On the other hand, all thermal X-ray emission mechanisms, used to explain the spectra of X-ray binaries till 100 KeV, like black body or Comptonization of low energy photons by energetic electrons, show an exponential cutoff after few keV decades. Nevertheless at the end of the Eighties some TeV Cherenkov experiments claimed the observation of photons coming from X-ray binaries. Microquasars, a subclass of high mass X-ray binaries, are now well established TeV sources and some of them have been recognized to be detected by EGRET. Moreover low mass X-ray binaries and black hole candidates have shown a non thermal hard tail, shaped like a power law that could extend at higher energies. Recently the MAGIC collaboration has presented evidence of TeV emission from Cyg X-1.

The third EGRET Catalogue of High-Energy Gamma-ray sources [23] contains 271 sources emitting radiation above 100 MeV. The catalogue includes a solar flare, the Large Magellanic Cloud, five pulsars, one probable radio galaxy (Cen A), and 66 high-confidence identification blazars plus 27 lower confidence potential blazars. No binaries were present in the catalogue (even if Vestrand et al. in 1997 [24] reported a signal in Cen X-3 at $5\sigma$ level of significance in October 1994).

Recently, X-ray emission has been detected by INTEGRAL from the two binaries IGR J6393-4643 [25] and IGR J16316-4028 [26] which have been associated to the EGRET unidentified sources: 3EG J1639-4702 and 3EG J1631-4033 respectively, and, in addition, the HESS collaboration claimed TeV emission from 3EG J1639-4702 [27].
On the other hand INTEGRAL and BeppoSAX observations have discovered that in the hard part of the spectrum of a large number of low mass X-ray binaries (e.g. Cyg X-1, GS 182-24, 4U 1636-53, Sco X-1) there is an excess which is well modelled by a power law (see e.g. reference [28]. The appearance of this component could suggest the presence of no-thermal emission mechanisms around these sources, as remarked by the observation of a jet in Cyg X-1 [29].

Early generation of Cherenkov telescopes at the end of Eighties and at the beginning of Nineties reported observations of gamma ray emission at the TeV energies for a group of binaries such as Cen X-3, SMC X-1, LMC X-4, Vela X-1, 4U 1145-619, AE aquarii (CV), and Her X-1 (see reference [30] and references therein). However, none of them has been confirmed by the new generation of IACT (Imaging atmospheric Cherenkov telescope) up to now. The reason of this defeat could be the bad quality of those observations or the fact that in those sources γ-ray is emitted in episodic events. This second hypothesis is not completely unreasonable if we consider the bursting and transient nature of the radiation of binaries [31].

On the other hand, more recently, a 3.9 days orbital modulation in the TeV flux from the X-ray binary LS 5039 has been reported by the HESS collaboration [32]. The observed modulation revealed, for the first time, that regions, where particles are accelerated up to $10^{12}$ eV, could exist around binary systems. Other evidence for variability at TeV energies has been recently unveiled in a similar type of binary system LS I +61 303 [33] and a TeV flare detected in coincidence with INTEGRAL and Swift from the direction of Cyg X-1 has been reported by the MAGIC collaboration [34]. All these facts suggest that binaries host extremely high energetic events in which relativistic particles could produce large amounts of gamma radiation.

Emission of high-energy gamma radiation is generally explained by two main physical mechanisms: hadronic interactions or inverse Compton scattering. In the hadronic scenarios relativistic protons are accelerated in jets or in shock regions, then, colliding and interacting with hadrons present in the interstellar medium, produce $\pi_0$ that decay in γ-ray radiation. The second scenario assumes a leptonic origin of gamma rays. In this case relativistic electrons, interacting with low energy photons, scatter the last ones to higher energies. Both scenarios need a physical mechanism to produce relativistic particles. Jets are generally invoked to explain high-energy radiation from AGNs, whereas shocks are used to justify emission in SNRs. In a different scale, both mechanisms could be present in binary systems. Shocks could form into the supersonic wind of the O star driven by jets appearing from the inner accretion disk around the compact object. Evidence of a relativistic jet has recently been reported for Cyg X-1 [29] in VLBA images.

With the advent of GLAST and AGILE and the new generation of Cherenkov telescopes such as HESS, MAGIC, VERITAS, and, in the future, the European CTA and the American AGIS, we will have the opportunity to extend the search for emission from X-ray binaries at very high energies to understand the still unexplored underlying physics.

At the moment only four binaries are recognized TeV emitters (PSR B 1259-63 [35], LS 5039, LS I +61 303, Cyg X-1). Cyg X-1 has been detected just in one episodic flare and it is unknown whether this emission has a cycle or not. The other sources show correlation between TeV flux and orbital period (for a wide exposition see the paper by Torres [36]) giving some constraints on the emission region. Present emission models usually propose two main different scenarios: Gamma rays come from region inside the pulsar wind zone or they are produced by accelerated particle in the wind collision shocks. On the other hand these models are studied in the contest of isolated object ignoring the role of the interaction with the companion. Future TeV instruments (CTA, AGIS) plan to improve the TeV sensitivity of an order of magnitude compared to the present IACTs. These could permit on one hand to significantly enlarge the source sample and on the other hand to perform short-timescales gamma-ray observations. The improved angular resolution could reveal essential in order to study the geometry of the jets giving information if the Gamma-rays are produced close the compact object, or farthest in the interaction whit the ISM.

The physics of binaries as TeV gamma ray emitter is largely still to be written. At the moment the phenomenology is poorly known and understood. New large arrays of Cherenkov telescopes will surely open new scenarios on this class of source that could still reserve some surprise.

## 2.5 Magnetars

Magnetars are a small class of isolated X-ray pulsars, thought to have the highest magnetic fields known to date: larger than $\sim 10^{14}$ - $10^{15}$ G, i.e. much larger than the quantum critical value $B_{QED}$ = $m^2c^3/\hbar\, e^-$ = 4.4 x $10^{13}$ G at which the energy between Landau levels of electrons equals their rest mass [37,38]. Magnetars have attracted increasing attention in the last decade, being extremely interesting objects, both from the physical and astronomical point of view. They allow us to observe and study several phenomena taking place in magnetic field conditions not available elsewhere (see, e.g., [39]). Their astrophysical importance is due to the fact that they broadened our view on how neutron stars are formed and evolve. Together with other new classes of neutron stars observed through the whole electromagnetic spectrum, they indicate that the classical radio pulsars discovered 40 years ago are just one of the diverse manifestations of neutron stars.

The class of magnetars comprises the Anomalous X-ray Pulsars (AXPs) and the Soft Gamma-ray Repeaters (SGRs), observationally very similar in many respects: a spin period in 2–12 s range, large period derivatives ($10^{-13}$ - $10^{-10}$ s s$^{-1}$), unpredictable bursting activity on different timescales (from ms to hundreds of seconds) and luminosities of $10^{38}$ - $10^{46}$ erg s$^{-1}$ (see [40] for a recent review). However, what differentiates these two classes of objects is, at present, unclear. Moreover, in the P-Pdot plane (see Figure 2.5.1), AXPs and SGRs lie in a region where other classes of isolated neutron stars are present. Again, whether or not the latter classes are linked to AXPs and SGRs is still under debate. Both AXPs and SGRs are associated to young structures: Supernova remnants for 3 AXPs and massive stellar clusters for 3 SGRs.

The magnetar model is originally founded on two observational facts: firstly, the rotational energy loss inferred from the SGR and AXP spin-down is insufficient to power their persistent X–ray luminosity of $\sim 10^{34}$ - $10^{36}$ erg s$^{-1}$; secondly, they apparently lack for companion stars which could provide the mass to power the X–ray emission through accretion. Through the years, several indirect indications have been collected all suggesting a high magnetic field for these sources [41,42]. Within the magnetar model the short ($t_b$<0.1s, $L_X$ < $10^{41}$ erg s$^{-1}$) bursts are produced by the propagation of Alfvèn waves in the magnetosphere, triggered by fractures in the crust. Rare (one every 10yr for the whole sample of SGRs) and energetic ($L_X$ up to $\sim 10^{47}$ erg s$^{-1}$) events, called giant flares, may occur when the fractures involve a large fraction of the NS. Both kinds of events are expected to induce a flux enhancement at all wavelengths.

Until not long ago AXPs and SGRs were thought as persistent and stable X-ray sources. Only in 2003 the first transient AXP was discovered by RXTE, namely XTE J1810-197, which displayed a factor of >100 flux enhancement with respect to the unpulsed pre-outburst quiescent luminosity level as seen by ROSAT and Einstein missions ($\sim 10^{33}$ erg s$^{-1}$; [43-47]). Even more surprising was the discovery of a highly variable pulsed radio emission that followed the XTE J1810-197's outburst [48,49], never observed before in any other magnetar [50]. Moreover, the transient nature of this AXP was the first hint that a relatively large number of members of this class is not discovered yet, and suggested that others would manifest themself in the future through a phenomenology (outburst) similar to that displayed by XTE J1810–197.

Indeed, after this first discovery, three other transient mangetars have been detected showing similar outbursts, CXOU J164710.2-455216 [51,52], SGR0501+4516, and 1E 1547.0-5408. The latter AXP was first suggested as a candidate magnetar in the supernova remnant G327.24-0.13 through X-ray observations [53], and subsequently confirmed as a radio transient magnetar through the discovery of radio pulsations [54,55] and of an X-ray outburst [56]. All these findings make magnetars not that dissimilar form pulsar, at least in terms of broadband energy emission from radio to GeV.

Finally, the discovery with the INTEGRAL satellite of persistent hard X-ray tails extending to ~250 keV in AXPs (no signature for a cut-off above 250keV has been yet unambiguously detected) came as a surprise, considering their soft spectra below 10 keV [57-61]. The hard spectra of magnetars together with their location in SNRs and stellar clusters made them more promising targets for higher energy instruments (above the MeV-GeV band).

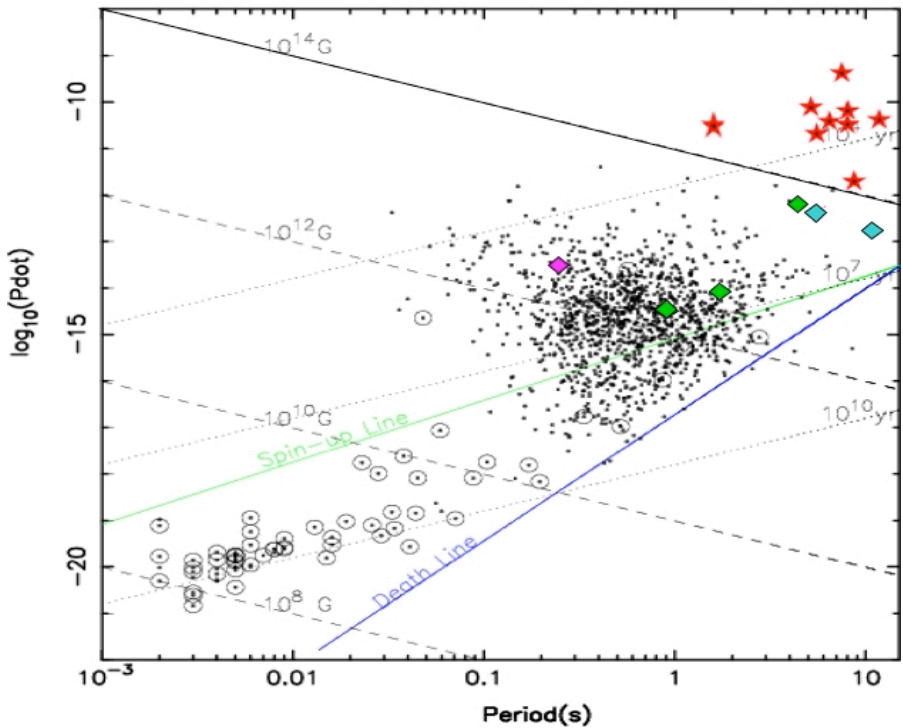

*Fig. 2.5.1.* P–Pdot plane for known isolated radio pulsars (dots), radio pulsars in binary systems (circled dots), together with the known AXPs and SGRs (filled stars). Even though AXPs and SGRs are clearly separated by the majority of radio pulsars, there are a number of them with similar P and Pdot, as well as 2 dim thermal neutron stars and RRATs (Rotating RAdio Transients; filled diamonds).

On shorter timescales of variability, the flaring activity of AXPs and SGRs is one of the most energetic, unpredictable and challenging phenomenon in Galactic high energy astrophysics. How and when these events occur, and the possible relation among their different manifestations, are still unsettled issues. The prompt study of timing and spectral characteristic of magnetar candidates, during and in the aftermath of flaring events, is of importance to shed light on the physical mechanisms responsible for their occurrence. The most powerful flares, called Giant Flares, are by far the most energetic (~$10^{44}$ - $10^{47}$ erg s-1) Galactic events currently known, second only to Supernova explosions. Only three events have been recorded in three decades of monitoring of the high energy sky, the last and more energetic one on 2004 December 27th [62]. They are characterised by a very luminous hard peak which has been observed until ~ 20MeV without any cut-off [63], lasting about a second and with a luminosity of about several $10^{47}$ erg s$^{-1}$ . Following the strong peak, which saturated every satellite in orbit around the Earth, the flare decays rapidly into a softer tail (observed until ~ 200 keV) lasting hundreds of second. Recent theoretical calculations suggest the presence of emission up to GeV for these rare events.

The future TeV experiments are thought to represent a new important window in the study and understanding of both the persistent and flaring emission of the enigmatic class of magnetars.

**References.**
[1] Bell, A.R., 1978, MNRAS, 182, 147
[2] Reynolds, S.P. 2008, ARA&A, 46, 89
[3] Völk, H. J., Berezhko, E. G., and Ksenofontov, L. T., 2005, A&A, 433, 229
[4] Blasi, P., 2005, Mod. Phys. Lett. A, 20, 3055
[5] Hillas, A.M., 2005, J. Phys. G, 31, 95
[6] Aharonian, F., et al. (HESS Coll.), 2007, A&A, 464, 235
[7] Aharonian, F., et al. (HESS Coll.), 2007a, ApJ, 661, 236
[8] Tanaka, T., et al., Preprint arXiv:0806.1490
[9] Amato, E., and Blasi, P., 2006, MNRAS, 371, 1251
[10] Morlino, G., Amato, E., and Blasi, P., 2009, MNRAS, 392, 240
[11] Gabici, S., Aharonian, F., and Blasi, P., 2007, Ap&SS, 309, 365
[12] Gabici, S., Aharonian, F., and Casanova, S., Preprint arXiv:0901.4549
[13] Aharonian, F. et al., 2006 A&A, 448, L43
[14] Aharonian, F. et al. [HESS Collaboration], 2005, A&A 435, L17
[15] Amato, E., and Arons, J., 2006, ApJ, 653, 325
[16] Amato, E., Guetta, D., and Blasi, P., 2003, A&A, 402, 827
[17] Aharonian, F. et al., 2005a, Science 307, 1938
[18] Aharonian, F. et al., 2006a, A&A, 442, L25
[19] Aharonian, F. et al., 2007, A&A, 472, 489
[20] Kaspi, V. M., Roberts, M. S. E., & Harding, A. K. 2006, Compact stellar X-ray sources, 279
[21] Thompson, D. J., Bailes, M., Bertsch, D. L., et al. 1999, ApJ, 516, 297
[22] Aliu, E., et al., 2008, Science, 322, 1221
[23] Hartman, R. C., et al., 1999, ApJS, 123, 79
[24] Vestrand, W. T., et al. 1997 ApJ, 483, 49
[25] Malizia, A. et al., 2004, ATel, 227, 1
[26] Rodriguez, J., and Goldwurm, A., 2003, Atel, 201,1
[27] Aharonian, F., et al., 2005, Science,307,1938
[28] D'amico, F., et al, 2001, ApJ, 547L, 147
[29] Stirling, A. M., 2001, MNRAS, 327, 1273
[30] Kifune, T., 1996, Space Science Review, 75, 31
[31] Psaltis D., 2004, in "Compact Stellar X-ray Sources", eds. W.H.G. Lewin and M. van der Klis, Cambridge Univerity Press
[32] Aharonian, F., et al., 2006, A&A, 460, 743
[33] Albert, J., et al., 2005, Science, 309, 746
[34] Albert, J., et al., 2007, ApJ, 665, 51
[35] Aharonian, F., et al., 2005, A&A, 442, 1
[36] Torres D. F. arXiv:0811.1137v2
[37] Duncan, R.C. & Thompson, C., 1992, ApJ, 392, L9
[38] Thompson, C. & Duncan, R.C., 1993, ApJ, 408, 194,
[39] Harding, A.K. & Lai, D, 2006, Rep Prog Phys, 69, 2631
[40] Mereghetti, S. 2008, A&A Rev., 15, 225
[41] Stella, L., Dall'Osso, S., et al. 2005, ApJ, 634, L165
[42] Vietri, M., Stella, L., et al. 2007, ApJ, 661, 1089
[43] Ibrahim, A.I., Markwardt, C., et al. 2004, ApJ, 609, L21
[44] Gotthelf, E.V., Halpern, J.P. 2004, ApJ, 605, 368
[45] Rea, N., Testa, V., et al. 2004, A&A, 425, L5
[46] Israel, G.L., Rea, N., et al. 2004, ApJ, 603, L97
[47] Bernardini, F., Israel, G.L. et al. 2009, A&A, in press, arXiv:0901.2241
[48] Camilo, F., Ransom, S.M, et al. 2006, Nature, 442, 892
[49] Helfand, D.J., Chatterjee, S., et al. 2007, ApJ, 662, 1198
[50] Burgay, M., Rea, N., et al. 2006, MNRAS, 372, 410
[51] Israel, G.L., Campana, S., et al. 2007, ApJ, 664, 448

[52] Muno, M.P., Gaensler, B.M., et al. 2007, MNRAS, 378, L44
[53] Gelfand, J.D., Gaensler, B.M., et al. 2007, ApJ, 667, 1111
[54] Camilo, F., Ransom, S.M., et al. 2007, ApJ, 666, L93
[55] Camilo, F., Reynolds, J., et al. 2008, ApJ, 679, 681
[56] Halpern, J.P., Gotthelf, E.V., et al. 2008, ApJ, 676, 1178
[57] Kuiper, L., Hermsen, W., et al. 2006, ApJ, 645, 556
[58] Kuiper, L., Hermsen, et al. 2004, ApJ, 613, 1173
[59] Molkov, S.V., Cherepashchuk, A.M., et al. 2004, Astronomy Letters, 30, 534
[60] Götz, D., Mereghetti, S., et al. 2006, A&A, 449, L31
[61] den Hartog, P.R., Hermsen, W., et al. 2006, A&A, 451, 587
[62] Hurley, K., Boggs, S.E., et al. 2005, Nature, 434, 1098
[63] Boggs, S., Hurley, K., et al. 2005, GCN, 2936

# 3 Extragalactic Astrophysics.

## 3.1 Blazars and Radio Galaxies.

### 3.1.1 Status of observations.

The present list of extragalactic TeV sources comprises 27 AGNs (24 blazars and 3 radiogalaxies)[9]. While only 2 of them (Mkn 421 and Mkn 501) were known in 1998 and 7 in 2003, in the last five years the number increased impressively, thanks to the new telescopes (mainly HESS, MAGIC and VERITAS). The distance of the sources also increased considerably: while the first confirmed TeV AGNs were rather local ($z < 0.05$), some of the sources recently discovered lies at redshift larger then 0.2 (and several distant sources have uncertain and even unknown redshift). The present record belong to the quasar 3C279, located at $z = 0.536$.

The large majority of these objects are blazars, radio-loud AGNs characterized by relativistic jets pointing close to the line of sight. The overall spectral energy distribution of blazars, extending from the radio band to TeV energies, shows two broad components that, for all TeV sources but 3C279, peaks in the optical-UV-X-rays band and in the GeV-TeV band. The level and the position of the two bumps is variable in time, also on very short timescales. Most of the TeV emitting blazars belongs to the subclass of the high energy peaked BL Lac objects (HBL), blazars in which the low energy component of the Spectral Energy Distribution (SED) peaks in the UV-X-ray band. Only few low energy peaked BL Lac objects (LBL, peaking in the optical-IR band) and only one quasar have been detected. HBL are the most powerful sources in the TeV band, since the high-energy component peaks just in this spectral region. LBL (and quasars), instead, present less powerful TeV tails. This fact, together with the average larger distance (determining also an increasing role of the absorption through the interaction with the extragalactic background light, see below) translates into a lower TeV fluxes for these sources.

However, an important caveat in considering the relative populations of TeV blazars is that, due to the very limited field of view, present instruments do not allow a survey of the sky. Instead, telescopes are pointed towards the best candidates in term of expected flux that, in fact, are the HBLs. In this respect, a catalogue of TeV blazar candidates often used to select the targets is contained in [1].

Therefore, the present knowledge of the extragalactic TeV is certainly biased. The recent discovery of the quasar 3C279 by MAGIC [2] demonstrates that also non-BL Lac blazars can be TeV emitters. These considerations highlight the importance of surveys and the decrease of the low energy threshold (to minimize the role of cosmic absorption for distant sources) for future studies in the TeV band. A survey at low energies (< 100 GeV) will be performed in the next years by the gamma-ray satellite Fermi, which is expected to provide several new TeV candidates.

### 3.1.2 Astrophysical framework.

While there are no doubts that the low energy emission of blazars is synchrotron radiation from energetic electrons in the relativistic jet, more debated is the origin of the TeV emission. The most popular scenario assumes that high-energy photons originate in the inverse Compton scattering between soft ambient photons and the same relativistic electrons responsible for the low-energy component. In particular for BL Lacs, due to the absence of important sources of soft photons outside the jet, it is widely assumed that soft photons are dominated by the synchrotron photons (Synchrotron–Self Compton, SSC, model, [3]).

---

[9] The full list and additional information can be found at: http://www.mppmu.mpg.de/~rwagner/sources

Though the SSC scenario is supported by a large number of observational clues, some recent observations are difficult to accomodate in this scheme. One of the most convincing evidence supporting the SSC model is the observed strict correlation between the variations in the X-ray and in the TeV band (e.g. [4]). Since in the SSC model, the emission in the two bands is produced by the same electrons, a strict correlation is expected. However, there are (rare) exceptions, the most famous being the so-called "orphan" TeV flares, not accompanied by variations in the X-ray band [5].

A direct generalization of the SSC model is a scenario in which other radiation field can act as target for the IC scattering. These classes of model (external Compton models, e.g. [6]) are usually applied to interpret the phenomenology of the powerful blazars, associated to flat spectrum radio quasars, displaying luminous emission lines and blue bump (the direct emission from the accretion disk). Among the alternatives to the inverse Compton scenarios discussed in literature, the most promising is the synchro–proton model [7], considering the direct synchrotron emission by ultra-relativistic protons ($E > 10^{19}$ eV) co-accelerated with the electrons.

### 3.1.3 *New problems and open issues.*

The recent important advances in the instrumental performances have raised new problems and questions of great astrophysical relevance. Among the most important we recall:

- **Large bulk Lorentz factors:** for several sources, the SSC model requires extremely large bulk Lorentz factor of the flow ($\Gamma = 50-100$), at odds with the small (even subluminal) velocity inferred with the VLBI. A possible solution invokes the presence of a structure in the jet, with a fast spine surrounded by a slower layer. The radiative interplay between the two regions is increased by the relative velocity and helps to lower the required $\Gamma$ [8]. Moreover, with the same structure it is possible to explain the observed TeV emission of radiogalaxies [9].
- **Ultra-fast variability:** short-living flaring episodes have been observed since the first TeV observations (e.g. the 20 minute flare in Mkn 421 observed by [10]). A recent and somewhat extreme example of short variability timescale is that of the TeV emission of PKS 2155-304, in which well-resolved bursts varying on time scales of 200 seconds have been observed [11]. Similar timescales have been observed for variation in Mkn 501 [12]. These observations raise important problems: what is the minimum timescale of flares? How common are these episodes? Where and how this rapidly-varying emission is produced? It is now clear that such small timescales cannot be related to the light crossing time of the central supermassive black hole, as commonly assumed before [13]. In this context there is even the possibility that flares as short as few seconds, with luminosities as large as $10^{47-48}$ erg s$^{-1}$ are produced [14].
- **Hard TeV spectra:** even considering the lowest allowed level of the Extragalactic Background Light (EBL), some HBL present rather hard spectra, really challenging the standard SSC model, in which the decrease of the scattering cross section, naturally leads to rather soft TeV spectra. Although alternatives have been discussed [15, 16, 17], we do not have a clear solution.
- **TeV quasars:** In the general framework depicted above, Flat Spectrum Radio Quasars (SRQ) are not expected to be powerful TeV emitters. In general, emitting particles have lower energies than those in BL Lac objects. Moreover, the high-energy peak is very likely dominated by the inverse-Compton emission off the ambient UV photons, determining a severe decrease of the emission above few tens of GeV, both for the reduction of the cross section and the absorption of gamma rays through the interaction with the ambient radiation field. However, the recent detection of 3C279 indicates that also quasars are, to some extent, TeV emitters. Although the detection of 3C279 can be located in the standard scenario [2], problems (in particular the role of the intrinsic absorption) are starting to emerge, requiring modification of the standard scenario (e.g. emission from regions of the jet far from the black hole) or even possibly indicating drastic changes of current ideas (e.g. hadronic emission, [18]). Note that if quasars can emit TeV photons they would become important probes for determining the EBL at UV-optical frequencies at relatively large redshift (where, instead, HBL are rare).

- **TeV radiogalaxies:** Radiogalaxies could be another interesting class of TeV sources. Presently only M87 and Centaurus A are confirmed source [19, 20], and 3C66B has possibly been detected by MAGIC [21]. TeV emission has been successfully explained as emission by misaligned structured jet (misaligned blazars, e.g. [9]), but also other possibilities have been considered (emission from particles accelerated close to the black hole horizon [22] or the jet at large scale [23]).

## 3.2 Extragalactic Background Light and the gamma-ray horizon.

Besides their astrophysical relevance, blazars are also interesting in view of the possible use as beacons to characterize the EBL through its interaction with gamma-ray photons.

The idea behind the use of blazars as probe of the EBL is rather clear: gamma-rays, in their travel from the source to the Earth can interact with the EBL photons through the pair production process $\gamma + \gamma \rightarrow e^+ + e^-$. Due to the characteristics of the the pair production cross section it is possible to assume that photons with a specific energy $E$ are basically selectively absorbed through the interaction with a soft photons of frequency $\nu = 2 \times 10^{15} (E/100\text{GeV})^{-1}$ Hz.

Ideally, comparing the observed spectrum of a blazars with the intrinsic one, one should be able to derive the intervening absorption and, in turn, characterize the spectrum and the level of the EBL. Using blazars located at different redshift one could even derive the evolution of the EBL with cosmic time.

The study of the EBL is rather interesting, since it contains the record of the star formation history of the Universe. Therefore, fundamental open problems, such as the evolution of the star formation and the role of the first stars (Pop III) could be addressed (e.g.[24]). However, severe observational problems (related to the presence of strong foregrounds from interplanetary dust) prevent direct measurements [25]. Given the large uncertainty in the measurements one has to resort to models to calculate the expected EBL and optical depth to gamma rays [26, 27, 28] (but see [29] for a different approach).

Therefore, although "indirect", the use of blazars is fundamental to derive constraints on the level of the cosmic background at IR, optical and UV frequencies. The interest for this topic is grown in the last decade, after the advent of sensitive Cherenkov telescopes, since absorption is particularly important in the VHE band.

Unfortunately, this simple approach is difficult in practice because of the poor knowledge of the intrinsic spectrum of blazars. An approach used to by-pass this problem is to assume some limiting hardness for the intrinsic spectrum. In practice, one does not expect that spectra with slopes harder than $\Gamma = 1.5$ (in photons) can be produced by standard processes. In this way, a maximum level of the EBL can be derived. This approach has been already applied using the data of nearby blazars (especially Mkn 501, for which photons with energies larger then 10 TeV have been detected, [30]). However, only in recent years rather interesting results have been obtained using blazars at larger distances, especially the BL Lac 1101-232 (z=0.182, [31]) and the quasar 3C279 (z=0.536, [2]). In both cases only rather low level of the EBL, close minimum to the level expected from galaxy counts [29], are allowed if the limit $\Gamma = 1.5$ is assumed. Criticisms have been raised to these results, mainly motivated by possible effects (efficient particle acceleration, intrinsic absorption of gamma-rays) producing intrinsic spectra harder than $\Gamma = 1.5$.

The improvement of these studies requires the extention of the number (and type) of sources. The final goal is to have a number of sources large enough to allow not only to infer the level (and the spectrum) of the EBL, but also to investigate the dependence of the derived EBL with redshift.

## *3.3 Star-forming galaxies.*

Diffuse gamma-ray emission from *pp* interactions of Cosmic Ray (CR) nuclei with target Inter-Stellar Matter (ISM) and photons makes up about 90% of the luminosity above 100 MeV of the Milky Way (Strong et al. 2000).

However, the corresponding VHE (>100 GeV) flux from a galaxy like the Milky Way located 1 Mpc away would be well below current IACT sensitivities. Indeed, among non-AGN-hosting galaxies only the LMC has so far been detected in gamma-rays, and only in the softer EGRET range (<100 MeV; Sreekumar et al. 1992).

However, high star-formation and supernova (SN) rates in starburst galaxies (SBGs) enhance the energy density of energetic nonthermal particles—mostly electrons and protons—which are Fermi-accelerated in the sites of SN remnants. Coulomb, synchrotron and Compton energy losses by the electrons, and the decay of pions following their production in energetic proton interactions with the ambient gas, result in emission over the full electromagnetic spectrum, from radio to high-energy gamma-ray. The higher level of emission in SBGs—compared to that in 'normal' galaxies—motivates consideration of the detection of nearby SBGs by current and future IACTs. When VHE gamma-ray emission is detected, important additional insight will be gained on the origin and propagation mode of energetic electrons and protons in SBGs.

However, only weak level of (isotropic) emission is expected, making only nearby or extremely powerful SBGs (i.e., M 82 and NGC 253; and Arp 220) the obvious targets for observations.

An approximate estimate of VHE emission from the hadronic process is obtained by an analytical calculation that takes into account the target gas distribution in the galaxy, the synchrotron radio emission as a calibrator of the number density of (primary plus secondary) electrons, computes the magnetic field by assuming equipartition of the field and particle energy densities.

A more accurate estimate is provided by a detailed numerical treatment based on a convection-diffusion model for energetic electron and proton propagation and energy losses, where all relevant hadronic and leptonic processes are considered, gauged by the measured synchrotron radio emission from the inner disk region. An initial particle spectrum, injected in the starburst region, is evolved through all the relevant leptonic and hadronic interactions as the particles diffuse and convect out of the acceleration region into the outer disk (and halo).

The ensuing estimates of $f_{(\geq 100\ GeV)}$ are: $\sim 2 \times 10^{-12}$ cm$^{-2}$ s$^{-1}$ for M 82 [34, 35], $\sim 1 \times 10^{-12}$ cm$^{-2}$ s$^{-1}$ for NGC 253 [36, 37], and $\sim 5 \times 10^{-12}$ cm$^{-2}$ s$^{-1}$ for Arp 220 [38].

There are no measurements of the VHE emission from M 82: the flux predicted by Persic et al. [34] could be detectable with deep exposures by MAGIC II and VERITAS, whereas CTA could detect it in about 10 hrs of observation.

Measurements of VHE emission from NGC 253 have so far resulted only in upper limits obtained from H.E.S.S observations [39]: this limit is close to recent estimates by Domingo-Santamaria & Torres [37] and Rephaeli et al. [36]. A MAGIC observation of the merging starburst Arp 220 has only returned a loose upper limit to the VHE flux [40], unable to constrain the prediction by Torres [38].

Detection of the VHE emission associated with ongoing star formation in the universe is clearly one major lingering goal of VHE astrophysics.

## 3.4 Gamma-ray Bursts

High Energy (HE) or Very High Energy (VHE) observations in the MeV-GeV-TeV range of Gamma-Ray Bursts (GRBs) have been suggested to be powerful diagnostic for the emission processes and physical conditions of GRBs by many authors [41, 42, 43, 44, 45, 46].

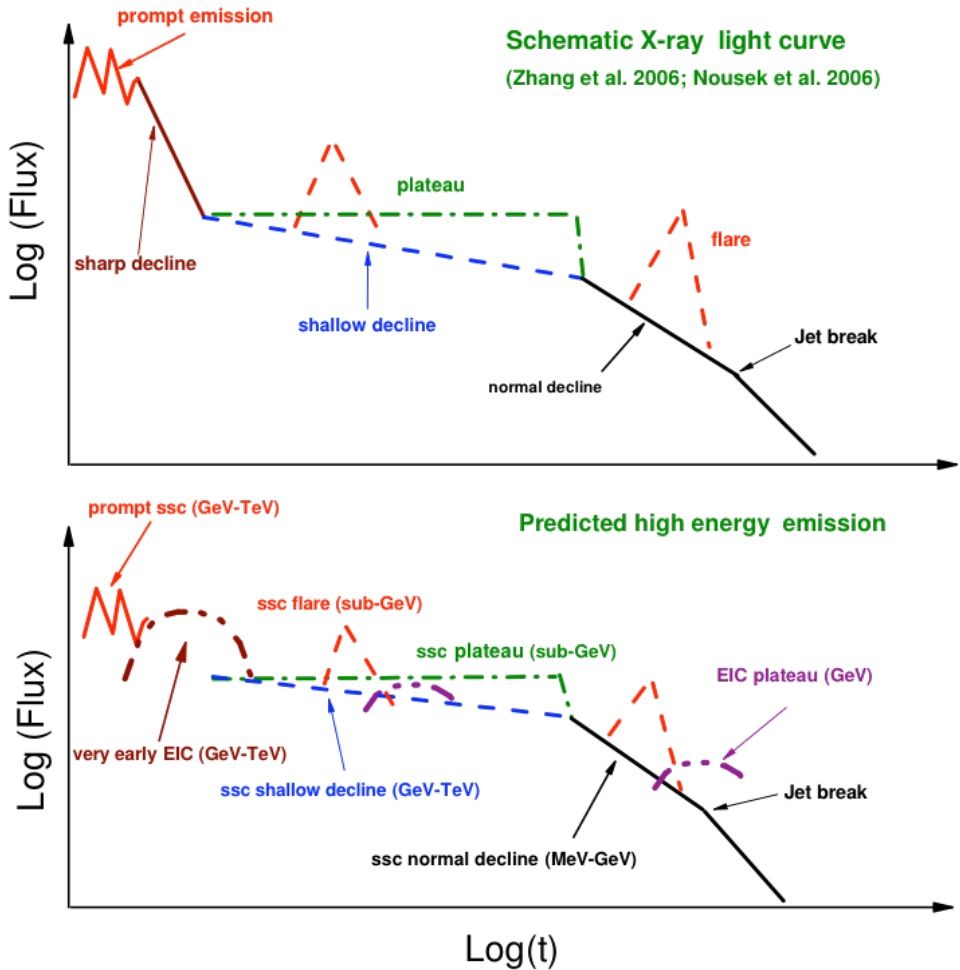

***Fig. 3.4.1:*** Schematic view of the typical soft-X-ray light curve for a GRB and possible high-energy processes shaping the higher energy light-curve (from [41]).

As a matter of fact, in spite of the many successes in the interpretation of the various phases of a GRB [47, 48], there are still fundamental unanswered questions involving essentially all aspects of the GRB phenomenology. In the past, several detections of GRBs were performed in the MeV-GeV range by the Energetic Gamma-Ray Experiment Telescope (EGRET) on-board the Compton Gamma-Ray Observatory (CGRO). HE emission was observed roughly coincident in time with the prompt emission (i.e. for GRB941017 [49]) or delayed and more naturally associated to the afterglow (i.e. GRB940217 [50]). More recently, MeV-GeV detections of GRBs have been also obtained by AGILE and FERMI. On the contrary, in spite of continuous attempts, so far no convincing detections at higher energies (~ TeV) have been obtained. Null detections have been reported by various Imaging Atmospheric Cherenkov Telescopes (IACTs) as HESS, VERITAS and, in particular, MAGIC which is characterized by the lowest energy threshold [51, 52, 53, 54, 55].

The universe at VHE is not transparent. Photons with energies above above tens of GeV can interact with the diffuse infrared photon background, or Extragalactic Background Light (EBL) [56]. The optical depth $\tau$ for a $\gamma \sim 100$ GeV photon at $z \sim 1$ can be as high as $\tau \sim 6$, depending on the specific model prediction. Observationally, for redshifts higher than about 0.2, the attenuation due to EBL interaction is uncertain, and the recent detection by the MAGIC telescope of the quasar 3C279 ($z = 0.54$) at energies above 300 GeV [57] could support more optimistic predictions with optical depth at $z \sim 1$ not far from unity. Most of the GRBs observed by IACTs are follow-up observations of alerts distributed by the Swift satellite [58] over the GRB Coordinates Network (GCN). The majority of Swift GRBs are at redshift substantially larger than one. Therefore, almost independently of the predicted flux at VHE from a GRB, unless the event is relatively nearby, the EBL attenuation makes very difficult to detect any emission.

Many good reviews about GRB interpretative scenarios are available. The leading picture is the so-called cosmological fireball model [59, 47, 60, 48]. In the fireball scenario the various observed phenomenology from a GRB are essentially due to an ultra-relativistic outflow (initial Lorentz $\Gamma$ factor between 100-1000 [61]) generated during the final collapse of a high mass star or the coalescence of a binary system made by compact objects. The former case is usually associated to the long duration GRBs, while the latter to the short duration class. Independently of details, the emission is supposed to happen in two distinct regions: (1) relatively nearby the central engine ($\sim 10^{13}$ cm) due to in-homogeneities in the outflow - the internal shocks, and (2) much further ($\sim 10^{16}$ cm), when the outflow interacts with the circum-burst medium - the external shocks. Synchrotron spectra intrinsically cover a wide energy range, and for GRBs the high-energy cutoff is mainly dependent on the Lorentz $\Gamma$ factor. With typical parameters of the prompt phase we could have photons up to a few GeV while in the early afterglow the photons extend up to energies of 10-100 MeV. Therefore already synchrotron emission can produce photons in the HE range as observed during the prompt phase in the past. On the contrary, GeV photons in the afterglow require some additional mechanism. The most important HE emission process for GRBs is likely Inverse Compton (IC) which is expected to be important in a large fraction of the micro-physical parameter space generating a synchrotron component [62, 63]. Depending on the specific conditions, it is possible to have multiple order IC scattering and at VHE energies the Klein-Nishina cross-section suppression has to be taken into account [64].

The cosmological fireball scenario offers a rich set of possibilities for an effective IC process. Synchrotron photons can be up-scattered by the same electrons which cooled by synchrotron emission (Synchrotron Self-Compton, SSC). SSC can occur both during the internal and the external shocks effectively producing a HE component following the time evolution of the underlying synchrotron component. However, the internal shocks (i.e. the prompt phase) is not, in general, the best place where to look for VHE emission due to the intense internal opacity preventing VHE photons to freely escape. A possible way to overcome this limitation is for GRBs characterized by a very high bulk Lorentz factor $\Gamma > 10^3$. It might also happen that electrons accelerated at the shock front up-scatter photons coming from different regions (External Inverse Compton, EIC), allowing many different possibilities (i.e. photons from reverse shock or delayed internal shocks interacting with electrons in the forward shock, etc. ). In this case the temporal relation between the synchrotron and the IC component gets more complicated since now in the co-moving frame photons are strongly beamed while electrons are still isotropic and the angular dependence of the IC scattering [64] can substantially delay (and lower, compared to the isotropic case) the total IC flux. For almost any reasonable spectrum VHE photons are intrinsically rare compared to lower energy photons and the higher the energy is the larger the required effective area should be. For a moderately bright (fluence $\sim 10^{-5}$ erg cm$^{-2}$ in the Swift -BAT range) GRB at $z\sim 1$ it is easy to estimate the number of photons FERMI can collect at MeV or GeV energies due to various processes [41, 46]. In general, apart from very bright events (like GRB080319B, the probability of which was roughly estimated to ~1/10 years [65]), FERMI should easily detect a few tens of photons at energies in the MeV range, while just a few photons can be detected in GeV range. On the contrary, in the tens of GeV range an IACT can again detect tens of photons. It appears clear that apart from the mere detection, for a reliable spectral analysis with the perspective to single out which emission process is efficient, the simultaneous availability of FERMI and an IACT is mandatory. Not only the statistics of detected photons is substantially improved, but the energy range leverage makes any analysis much more constraining. A low energy cut-off as low as 50 GeV seems to be a required feature to improve the capabilities of an IACT to detect emission from a GRB. The foreseen improvements of MAGIC, HESS and VERITAS seem all to be able to guarantee an effective area of several tens of square meters at a few tens of GeV.

**References:**

[1] Costamante L., Ghisellini G., 2002, A&A, 384, 56
[2] Albert J. et al. 2008, Science, 320, 1752
[3] Tavecchio F., Maraschi L., Ghisellini G., 1998, ApJ, 509, 608
[4] Fossati G., et al., 2008, ApJ, 677, 906
[5] Krawczynski H., et al., 2004, ApJ, 601, 151
[6] Sikora M., Begelman M. C., Rees M. J., 1994, ApJ, 421, 153
[7] Aharonian F. A., 2000, NewA, 5, 377
[8] Ghisellini G., Tavecchio F., Chiaberge M., 2005, A&A, 432, 401
[9] Tavecchio F., Ghisellini G., 2008, MNRAS, 385, L98
[10] Gaidos J. A. et al. 1996, Nature, 383, 319
[11] Aharonian F., et al., 2007, ApJ, 664, L71
[12] Albert J., et al., 2007, ApJ, 669, 862
[13] Begelman M. C., Fabian A. C., Rees M. J., 2008, MNRAS, 384, L19
[14] Ghisellini G., Tavecchio F., Bodo G., Celotti A., 2009, MNRAS, 393, L16
[15] Katarzynski K., Ghisellini G., Tavecchio F., Gracia J., Maraschi L., 2006, MNRAS, 368, L52
[16] Bottcher M., Dermer C. D., Finke J. D., 2008, ApJ, 679, L9
[17] Aharonian F. A., Khangulyan D., Costamante L., 2008, MNRAS, 387, 1206
[18] Bottcher M. et al. 2009, ApJ, submitted (arXiv:0810.4864)
[19] Aharonian F., et al., 2006, Sci, 314, 1424
[20] Albert J., et al., 2008, ApJ, 685, L23
[21] Aliu E. et al., 2009, ApJ 692, L29
[22] Neronov A., Aharonian F. A., 2007, ApJ, 671, 85
[23] Stawarz L., et al. 2006, MNRAS, 370, 981
[24] Raue M, Kneiske T., Mazin D., 2009, A&A, in press (arXiv:0806.2574)
[25] Hauser M. G., Dwek E., 2001, ARA&A, 39, 249
[26] Stecker F. W., Malkan M. A., Scully S. T., 2006, ApJ, 648, 774
[27] Kneiske T. M., Bretz T., Mannheim K., Hartmann D. H., 2004, A&A, 413, 807
[28] Primack J. R., Bullock J. S., Somerville R. S., 2005, AIPC, 745, 23
[29] Franceschini A., Rodighiero G., Vaccari M., 2008, A&A, 487, 837
[30] Stanev T., Franceschini A., 1998, ApJ, 494, L159
[31] Aharonian F., et al., 2006, Nature, 440, 1018
[32] Strong A.W. et al. 2000, ApJ, 537, 763
[33] Sreekumar P. et al. 1992, ApJ, 400, L67
[34] Persic M., Rephaeli Y. & Arieli Y. 2008, A&A, 486, 143
[35] De Cea del Pozo E., Torres D.F.&Rodriguez Marrero A.Y., submitted
[36] Rephaeli Y., Arieli Y. & Persic M. 2009, A&A, submitted
[37] Domingo-S. E. & Torres D. 2005, A&A, 444, 403
[38] Torres D.F. 2004, ApJ, 617, 966
[39] Aharonian F.A. et al. 2005c, A&A, 442, 177
[40] Albert J. et al. (the MAGIC collaboration) 2007, ApJ, 658, 245
[41] Y.-Z. Fan, and T. Piran, 2008, astro-ph/ 0805.2221
[42] A.D. Falcone, D.A. Williams, M.G. Baring et al., 2008, astro-ph/ 0810.0520
[43] F. Aharonian, J. Buckley, T. Kifune, G. Sinnis, 2008, Rep. Prog. Phys. 71
[44] C.D. Dermer, and C.L. Fryer, 2008, astro-ph/ 0809.3959
[45] A. Galli, and L. Piro, 2008, Astron. Astrophys. 489, 1073
[46] T. Le, and C.D. Dermer, 2008, astro-ph/ 0807.0355

[47] B. Zhang, and P. M´esz´aros, 2004, Int. J. Mod. Phys. A 19, 2385
[48] B. Zhang, 2007, Chin. J. Astron. Astrophys. 7, 1
[49] M.M. González, B.L. Dingus, Y. Kaneko et al.,2003, Nature 424,749
[50] K. Hurley, B.L. Dingus, R. Mukherrjee et al., 1994, Nature 372, 652
[51] J. Albert, E. Aliu, H. Anderhub et al., 2006, Astrophys. J. 641, 9
[52] D.C.Morris, J. Reeves, V. Pal'shin et al., 2007, Astrophys. J. 654, 413
[53] D. Bastieri, N. Galante, M. Garczarczyk et al., 2007, astro-ph/ 0709.1380
[54] N. Galante, D. Bastieri, M. Gaug et al., 2008, AIPC 1000, 125
[55] M. Garczarczyk, L.A. Antonelli, A. La Barbera et al., 2008, APIC, 1065, 342
[56] M.G. Hauser, and E. Dwek, 2001, Ann. Rev. Astron. Astrophys. 39, 249
[57] J. Albert, E. Aliu, H. Anderhub et al., 2008, Science 320, 1752
[58] N. Gehrels, G. Chincarini, P. Giommi et al., 2004, Astrophys. J. 611, 1005
[59] T. Piran, 1999,Phys. Rep. 314, 575
[60] P. Mèszàros, 2006, Rep. Prog. Phys. 69, 2259
[61] E.. Molinari, S. Vergani, D. Malesani et al., 2007, Astron. Astrophys. 469, 13
[62] A. Pe'er, E. Waxman, 2004, Astrophys. J. 613, 448
[63] A. Pe'er, E. Waxman, 2005, Astrophys. J. 633, 1018
[64] G.B. Rybiki, and A.P. Lightman Radiation Processes in Astrophysics, Wiley,New York (1979)
[65] J.L. Racusin, S.V. Karpov, M. Sokolowski et al., 2008, Nature 455, 183

# 4 New Physics and *Exotica*.

## *4.1 Dark Matter.*

Evidence for discrepancy of cosmological motions from the predictions of Newtonian dynamics based on visible matter, interpreted as due to the the presence of DM, are well established—from galaxy scales [1] to galaxy-cluster scales [2] to cosmological scales [3].

DM particle candidates should be weakly interacting with ordinary matter (and hence neutral). The theoretically favoured ones, which are heavier than the proton, are the weakly interacting massive particles (WIMPs). WIMPs should be long-lived enough to have survived from their decoupling from radiation in the early universe into the present epoch. Except for the neutrino, which is the only DM particle known to exist within the Standard Model of elementary particles (with a relic background number density of ~50 cm$^{-3}$ for each active neutrino species but which is too light (m$_v$ ≤ 1 eV) to contribute significantly to $\Omega_M$ given the current cosmological model, WIMP candidates have been proposed only within theoretical frameworks mainly motivated by extensions of the Standard Model of particle physics (e.g., the R-parity conserving supersymmetry [SUSY]). Among current WIMP candidates (see [4]), the neutralino, which is the lightest SUSY particle, is the most popular candidate. Its relic density is compatible with WMAP bounds (see [5]).

WIMPs could be detected directly, via elastic scattering with targets on Earth, or indirectly, by their self-annihilation products in high-density DM environments. DM annihilation can generate gamma-rays through several processes. Most distinctive are those that result in mono-energetic spectral lines, $\chi\chi \rightarrow \gamma\gamma$, $\chi\chi \rightarrow \gamma Z$ or $\chi\chi \rightarrow \gamma h$. However, in most models the processes only take place through loop diagrams; hence the cross sections for such final states are quite suppressed, and the lines are weak and experimentally challenging to observe. A continuum gamma-ray spectrum can also be produced through the fragmentation and cascades of most other annihilation products. The resulting spectral shape depends on the dominant annihilation modes [6], whereas the normalization depends on the WIMP's mass and velocity-averaged annihilation cross-section as well as on the DM density profile. The role of internal bremsstrahlung from the virtual particles that mediate the annihilation in the neutralino decay process has been recently quantified, and found to be able to significantly enhance the emission [7].

Once the astroparticle model has been chosen (e.g., [8]), the main uncertainties are of astrophysical nature. Superposed to any VHE emission from the decaying DM (cosmological, non-baryonic signal), galaxies can display a VHE emission from astrophysical sources associated with the visible matter distribution (astrophysical, baryonic signal). The ratio of the former to the latter is maximized in small, low-luminosity, low-Star formation rate galaxy. This is because the dark-to-visible mass ratio, as well as the central DM density, increases with decreasing luminosity [9].

Clearly, distance dilution of the signal opposes detection, so galaxies candidate for indirect DM detection should be chosen among nearby objects. In conclusion, the best observational targets for DM detection are the Milky Way dwarf spheroidal galaxies (e.g., Draco, Sculptor, Fornax, Carina, Sextans, Ursa Minor). A further issue, stemming from the $\rho^2$ dependence (as a result of annihilation) of the normalization integral of the gamma-ray emission, concerns the shape of the inner halo profile, i.e. whether the latter is cuspy or cored. Cuspy profiles are produced in cosmological N-body simulations of halo formation [10], whereas cored profiles are suggested by the measured rotation curves of disk galaxies [11] also in low-surface-brightness galaxies, where the local self-gravity of baryons is virtually negligible [12].

These considerations (and uncertainties) have been incorporated in detailed predictions of the gamma-ray flux expected from the annihilation of the neutralino pairs. Outlooks for VHE neutralino detection in Draco by current IACTs are not very promising: for a neutralino mass $m_\chi$=100 GeV and a variety of annihilation modes, and in the favourable case of a maximal (cuspy) inner halo profile, VHE detection (by MAGIC in 40 hrs observation) can occur if average value of the neutralino's cross section times velocity is $\langle\sigma v\rangle \geq 10^{-25}$ cm$^3$ s$^{-1}$, which is somewhat larger than the maximum value for a thermal relic with a density equal to the measured (cold) DM density (but may be fine for non-thermally generated relics) in the allowed SUSY parameter space. The prospects are better in the HE range (100 MeV-10 GeV): for a maximal (cuspy) halo, 1 yr of GLAST observation should be able to yield a detection if $m_\chi \leq 500$ GeV and $\langle\sigma v\rangle \sim 3\times 10^{-25}$ cm$^3$ s$^{-1}$ [6, 13].

No evidence of DM annihilation gamma-rays has been unambiguously claimed so far. An apparently extended signal from the direction of NGC253, originally claimed by the CANGAROO collaboration [14] and attributed to the halo of NGC 253 [15] as arising from a combination of astrophysical emission [16] and DM-annihilation emission [17], was later definitely interpreted as due to hardware malfunction [18]. MAGIC observations of the "dark" galaxies Draco dSph and Willman I (the latter currently being the most promising Galactic satellite candidate, at D=35 kpc and with M/L ~700), have only returned upper limits ($f_{(>140\ GeV)} < 10^{-11}$ cm$^{-2}$ s$^{-1}$ and $f_{(\geq 100\ GeV)} < 10^{-12}$ cm$^{-2}$ s$^{-1}$ respectively: [19, 20]).

Clearly, IACT (+ Fermi/ *LAT*, possibly) positive detections from the direction to known dark halos, all characterized by the same spectral signatures, would be seen as a powerful indication that neutralino decay has been detected—and hence the nature of DM has been unveiled. That would be breaking news!

## *4.2 Test of the Lorentz Invariance.*

The constancy of the speed of light, *c*, is one of the postulates of the theory of relativity; it has been extensively tested in the recent years against a possible dependence on the photon energy. Besides the general interest in verifying this fundamental postulate, several theories (see [21] for reviews) predict Lorentz-invariance violations (LIV) via modifications of the propagation of energetic particles, namely dispersive effects due to a non-trivial refractive index induced, for example, by Quantum Gravity (QG) effects.

The dependence of the speed of light on the photon energy *E* is usually parametrized as

$$c' = c \cdot \left[1 \pm \left(\frac{E}{E_{s1}}\right) \pm \left(\frac{E}{E_{s2}}\right)^2 \pm \ldots\right] \qquad (4.2.1)$$

The energy scales $E_{s1}$, $E_{s2}$ which parametrize the LIV in Eq.(4.2.1) are usually expressed in units of the Planck mass, $M_P \sim 1.22\times 10^{19}$ GeV (natural units with $h/2\pi = c = 1$ are employed throughout). If the linear term dominates, the above expression reduces to:

$$c' = c \cdot \left[1 \pm \left(\frac{E}{E_{s1}}\right)\right] \qquad (4.2.2)$$

A favored way to search for such a dispersion relation is to compare the arrival times of photons of different energies arriving on Earth from pulses of distant astrophysical sources [22,23]. The greatest sensitivities may be expected from sources with short pulses, at large distances or redshifts *z*, of photons observed over a large range of energies. In the past, studies have been made of emissions from pulsars, γ-ray bursts (GRBs) and active galactic nuclei (AGN) (see [21] for a review).

Very recently, the IACTs and the *Fermi*/LAT γ-ray telescope provided us with the opportunity to test this hypothesis at an unprecedented precision. *(i)* MAGIC [24] has published the results of the analysis of a giant flare of the Mrk 501 blazar, at $z$=0.034. A correlation has been observed (with a probability of 2.5% of coming from a statistical fluctuation) between the photon energy and the arrival time. The correlation is *subluminal*, i.e., the minus sign holds in the above (linear) velocity dispersion of light (higher energy photons are delayed), with a coefficient $dt/dE = (0.030 \pm 0.012)$ s/GeV. This would indicate a value $E_{s1} \sim M_P/30$. The data have been also analysed in the hypothesis of no effect, giving the limit (at 95% C.L.) $E_{s1} > 0.02 M_P$. *(ii)* H.E.S.S. has detected a giant flare of the blazar PKS2155, observing no evidence of correlation between the photon energy and the time of arrival [25]. This allows setting a limit (at 95% C.L.) $E_{s1} > 0.06 M_P$. *(iii)* Finally, with the observation of the GRB080916C [26] at a photometric $z$=4.35±0.15, the *Fermi*/LAT detector has set the limit (at 95\% C.L.) $E_{s1} > 0.11 M_P$. This GRB, however, presents a correlation between the photons' energy and arrival time; in particular the most energetic photon, at $E = 13.2^{+0.70}_{-1.52}$ GeV, has arrived at 16.54 s after the primary burst.

The figures of merit for the sensitivity to effects related to new physics are the energy span ΔE and the redshift $z$. Especially for ΔE new generation Cherenkov telescopes will substantially improve the panorama.

## *4.3 Anomalies in photon propagation: Axion-Like Particles (ALPs).*

The horizon of the observable Universe is expected to rapidly shrink in the very-high-energy (VHE) band above 100 GeV as the energy further increases. This arises because photons from distant sources scatter off background photons permeating the Universe, thereby disappearing into electron-positron pairs [27]. The corresponding cross section $\sigma_{\gamma\gamma}$ peaks where the VHE photon energy $E$ and the background photon energy ε are related by $\varepsilon \approx 0.5\,(E/\text{TeV})^{-1}$ eV. Therefore, for observations performed with Imaging Atmospheric Cherenkov Telescopes (IACTs), which probe the energy interval 0.1–100 TeV, the resulting cosmic opacity is dominated by the interaction with ultraviolet/optical/infrared diffuse background photons (frequency band $1.2 \times 10^3 - 1.2 \times 10^6$ GHz, corresponding to the wavelength range 0.25 μm - 250 μm), usually called Extragalactic Background Light (EBL), which is produced by the stellar population of galaxies during the whole history of the Universe. Neglecting evolutionary effects for simplicity, photon propagation is controlled by the photon mean free path $\lambda_\gamma(E)$ for $\gamma\gamma \rightarrow e^+e^-$, and so the observed photon spectrum $\Phi_{obs}(E,D)$ is related to the emitted one by

$$\Phi_{obs}(E,D) = \exp[-D/\lambda_\gamma(E)]\,\Phi_{em}(E) \qquad (4.3.1)$$

Within the energy range in question, $\lambda_\gamma(E)$ decreases like a power law from the Hubble radius ≈ 4.2 Gpc at ≈ 100 GeV to ≈ 1 Mpc at ≈ 100 TeV [8]. Thus, Eq.(4.3.1) entails that the observed flux is *exponentially* suppressed both at high energy and at large distances, so that sufficiently far-away sources become hardly visible in the VHE range and their observed spectrum should anyway be *much steeper* than the emitted one.

Yet, observations have *not* detected the behaviour predicted by Eq.(4.3.1). A first suggestion in this direction came with the H.E.S.S. discovery [29] of the blazars H 2356-309 ($z$=0.165) and 1ES 1101-232 ($z$=0.186) at $E\sim 1$ TeV. Stronger evidence comes from the MAGIC observation [30] of 3C 279 ($z$=0.536) at $E\sim 0.5$ TeV: in particular, the signal at $E<220$ GeV has similar statistical significance as that for 220 GeV $< E <$ 600 GeV (6.1σ and 5.1σ, respectively).

A suggested way out of this difficulty relies upon the modification of the standard Synchro-Self-Compton (SSC) emission mechanism. One option invokes strong relativistic shocks [31]. Another rests upon photon absorption inside the source [32]. While successful at substantially hardening the emission spectrum, all these attempts fail to explain why *only* for the most distant blazars does such a drastic departure from the SSC emission spectrum show up.

An alternative proposal – usually referred to as the DARMA scenario [33] – can be summarized as follows. Implicit in the previous considerations is the hypothesis that photons propagate in the standard way throughout cosmological distances. In the DARMA scenario it is instead supposed that photons can oscillate into a new very light spin-zero particle – named Axion-Like Parlicle (ALP) – and vice-versa in the presence of cosmic magnetic fields, whose existence has definitely been proved by AUGER observations [34]. Once ALPs are produced close enough to the source, they travel *unimpeded* throughout the Universe and can convert back to photons before reaching the Earth. Since ALPs do not undergo EBL absorption, the *effective* photon mean free path $\lambda_{\gamma\,eff}(E)$ gets *increased* so that the observed photons cross a distance in excess of $\lambda_\gamma(E)$. Correspondingly, Eq.(4.3.1) becomes

$$\Phi_{obs}(E,D) = \exp[-D/\lambda_{\gamma\,eff}(E)]\ \Phi_{em}(E) \qquad (4.3.2)$$

from which we see that even a *slight* increase in $\lambda_\gamma(E) \to \lambda_{\gamma eff}(E)$ gives rise to a *huge* enhancement of the observed flux. It turns out that the DARMA mechanism makes $\lambda_{\gamma eff}(E)$ shallower than $\lambda_\gamma(E)$ although it remains a decreasing function of *E*. So, the resulting observed spectrum is *much harder* than the one predicted by Eq.(4.3.2), thereby ensuring agreement with observations even for a *standard* SSC emission spectrum. As a bonus, a natural explanation emerges for the fact that only the most distant blazars would demand $\Phi_{em}(E)$ to substantially depart from the emission spectrum predicted by the SSC mechanism.

Remarkably enough, the key ingredient of the DARMA scenario – namely the existence of ALPs – is not an *ad hoc* assumption invented to solve the problem in question. Instead, very light ALPs turn out to be a generic prediction of many extensions of the Standard Model of elementary particle physics and have attracted considerable interest over the past few years. Besides than in four-dimensional models [35], they naturally arise in the context of compactified Kaluza-Klein theories [36] as well as in superstring theories [37]. Moreover, it has been argued that an ALP with mass *m* ~ $10^{-33}$ eV is a good candidate for the quintessential dark energy [38] which might trigger the present accelerated cosmic expansion.

Imagine now that a sizeable fraction of photons emitted by a blazar soon convert into ALPs. They propagate unaffected by the EBL and suppose that before reaching the Earth a substantial fraction of ALPs is converted back into photons. Assume further that this photon-ALP oscillation process is triggered by cosmic magnetic fields (CMFs) [34]. Owing to the notorious lack of information about their morphology, one usually supposes that CMFs have a domain-like structure [39]: i.e., ***B*** ought to be constant over a domain of size $L_{dom}$ equal to its coherence length with ***B*** randomly changing its direction from one domain to another but keeping approximately the same strength. It looks plausible to assume the coherence length in the range 1-10 Mpc. Correspondingly, the inferred strength lies in the range 0.3-1.0 nG [40] (below 100 GeV where EBL absorption is negligible the effect would merely be a dimming [41]).

The ultimate goal consists in the evaluation of the probability $P_{\gamma\to\gamma}(E,D)$ that a photon remains a photon after propagation from the source to us when allowance is made for photon-ALP oscillations as well as for photon absorption from the EBL. As a consequence, Eq.(4.3.2) gets replaced by

$$\Phi_{obs}(E,D) = P_{\gamma\to\gamma}(E,D)\ \Phi_{em}(E). \qquad (4.3.3)$$

One first solves exactly the beam propagation equation, arising from the relevant Lagrangian over a single domain and assuming that the EBL is described by the ``best-fit model'' of [42]. Starting with an unpolarized photon beam, one next propagates it by iterating the single-domain solution as many times as the number of domains crossed by the beam, taking each time a *random* value for the angle between ***B*** and a fixed overall fiducial direction. Such a procedure is repeated $10^4$ times and next the average over all these realizations of the propagation process is taken. It is found that ~13% of the photons arrive to the Earth for *E*=500 GeV, representing an enhancement by a factor of about 20 with respect to the expected flux without DARMA mechanism (the comparison is made with the above "best-fit model"). The same calculation gives a fraction of 76% for *E*=100 GeV (to be compared to 67% without DARMA mechanism) and a fraction of 3.4% for *E*=1 TeV (to be compared to 0.0045% without DARMA mechanism). These conclusions hold for ALP mass *m* << $10^{-10}$ eV and the inverse two-photon coupling M of the ALP has been taken for definiteness M≈ 4×$10^{11}$ GeV but practically nothing changes for $10^{11}$ GeV < M < $10^{13}$ GeV.

This prediction can be currently tested with the space-borne *Fermi* telescope as well as the ground-based IACTs (H.E.S.S., MAGIC, CANGAROO III, VERITAS) and extensive-air-shower arrays (ARGO-YBJ, MILAGRO).

## *4.4 Top-down mechanisms.*

Extragalactic γ-ray emission could originate in decays of exotic particles in the early Universe. The energy spectrum of this component should be different from the AGN contributions [43][44]. Bounds on long-lived relics have been derived using EGRET and COMPTEL observations of the diffuse γ-ray background [45]. Many models predict long-lived relics that may or may not be dark matter candidates. Long lifetimes for heavy relics, larger than the age of the Universe, may arise in models with symmetry breaking at short distances. Examples of such models are technibaryons in technicolor models or the lightest supersymmetric particle in an R-parity violating SUSY model.

## *4.5 Optical SETI.*

The most promising technique in the search for extraterrestrial intelligence through optical messengers (OSETI) is to look for intense optical photons pulses from star systems [46]. Howard et al. [47] noted that, with current laser technologies, 3 ns optical pulses could be broadcasted that would be detectable at a distance of 1000 light-years, outshining starlight from the host system by a factor of $10^4$. A number of dedicated projects are at work (see [47] and references therein).

The detection of such short, intense light pulses, requires fast photon detectors. While diffraction limited resolution is necessary for all potential transmitters in order to maintain a small beam width, the optical properties of the receiver are less critical. Basically one needs a large reflector to collect the largest possible number of photons, and fast photo-detectors to discriminate the signal pulse from the night sky background. Ground-based gamma-ray telescopes, being equipped with large mirrors and GHz electronics, are ideal instruments for this type of observation [48,49].

Among the present Cherenkov telescopes, MAGIC has the largest area and a 2~GHz DAQ, and is therefore the most suitable for this research. The MAGIC collaboration has designed a control system for OSETI Pulse trigger level hardware [50]. An optical laser pulse coming from the direction of a candidate stars appears as a point of origin in the camera.

Given the large size and the number of telescopes, the next generation of air Cherenkov telescopes will outperform MAGIC .

## *4.6 Energy threshold effects in absorption processes*

It has been suggested by several authors [51,52] (and references therein) that a powerful tool to investigate Planck-scale departures from Lorentz symmetry could be provided by certain types of energy thresholds in the pair production process $\gamma_{VHE}\, \gamma_{EBL} \rightarrow e^+e^-$ of gamma-rays from cosmological sources.

In the conventional description, the EBL give rise to strong absorption of TeV photons, limiting the γ-ray horizon. The standard process is $\gamma_{VHE}\, \gamma_{EBL} \rightarrow e^+e^-$ production, whose corresponding cross section can be calculated [53]. It turns out that, the threshold condition for this process is considerably affected by effects that violate Lorentz symmetry.

In a collision between a soft photon of energy ε and a high-energy photon of energy $E$, an electron-positron pair could be produced only if $E$ is greater than a certain threshold energy $E_{th}$, which depends on ε and $m_e^2$. Using a dispersion relation [22] of the form

$$m^2 \approx E^2 - \boldsymbol{p}^2 + \xi\, \boldsymbol{p}^2\, (E^n/E_p^n) \qquad (4.6.1)$$

with real ξ and $n$ integer (>0), one obtains, for $n$=1 and unmodified law of energy-momentum conservation, that for a given soft-photon energy ε, the process $\gamma\gamma \rightarrow e^+e^-$ is allowed only if $E$ is greater than a certain threshold energy $E_{th}$ which depends on ε and $m_e^2$:

$$E_{th}\, \varepsilon + \xi\, (E_{th}^3/8E_p) \approx m_e^2 \qquad (4.6.2)$$

The ξ→0 limit corresponds to the special-relativistic result $E_{th} = m_e^2/\varepsilon$. For ξ~1 and sufficiently small values of ε (and correspondingly large values of $E_{th}$) the Planck-scale correction cannot be ignored.

This provides an opportunity for pure-kinematics tests. As an example, a 10~TeV photon and a 0.03~eV photon can produce an electron-positron pair according to ordinary special-relativistic kinematics, but they cannot produce a $e^+e^-$ pair according to the dispersion relation in Eq.(4.6.1) with $n$=1 and ξ~ −1. The situation for positive ξ is somewhat different, because a positive ξ decreases the energy requirement for electron-positron pair production.

This kind of effects can be best seen considering TeV photons emitted by blazars, for which the infrared diffuse extragalactic background photons are potential targets for electron-positron pair production. According to [52], the sensitivities that we can expect to achieve with this type of analysis are less than a factor 100 away from Planck-scale. Of course one has to take into account the fact that there are at least three other factors thay are important in establishing the amount of absorption of TeV photons emitted by a given blazar: our knowledge of the type of signal emitted by the source, the distance of the blazar and, most importantly, the density of the infrared diffuse extragalactic background.

In the past few years the availability of observations of the relevant type has very significantly increased, and certainly within a few years dramatic improvements will occur. This means that this strategy of measurements and analysis will eventually take us at the Planck-scale sensitivity and beyond.

# References.

[1] van Albada T.S. et al. 1985, ApJ, 295, 305
[2] Sarazin C. 1986, Rev. Mod. Phys., 58, 1
[3] Spergel D.N. et al. 2003, ApJS, 148, 175
[4] Bertone G. et al. 2005 Phys. Rep. 405, 279
[5] Munoz C. 2004, Int.J.Mod.Phys. A, 19, 3093
[6] Bergstrom L. & Hooper D. 2006, Phys. Rev. D, 73, 063510
[7] Bringmann T., Bergström L. & Edsjo J. 2008, JHEP, 801, 49
[8] Bergstrom L. et al. 1998, Astropart. Phys., 9, 137
[9] Persic M. et al. 1996, MNRAS, 281, 27
[10] Navarro J.F. et al. 1997 ApJ, 490, 493
[11] Borriello A. & Salucci P. 2001, MNRAS, 323, 285
[12] De Bloek W.J.G. et al. 2001, ApJ, 552, L23
[13] Sanchez-Conde M.A. et al. 2007, Phys.Rev.D, 76, 12, 123509
[14] Itoh C. et al. 2002, A&A, 396, L1
[15] Itoh C. et al. 2003a, A&A, 402, 443
[16] Itoh C. et al. 2003b, ApJ, 584, L65
[17] Itoh C. et al. 2003c, ApJ, 596, 216
[18] Itoh C. et al. 2007, A&A, 462, 67
[19] Albert J. et al. (the MAGIC collaboration) 2008, ApJ, 679, 428
[20] Aliu E. et al. (the MAGIC collaboration) 2009, ApJ, in press
[21] Sarkar, S. 2002, Mod. Phys. Lett. A 17, 1025.
[22] Amelino-Camelia, G., et al. 1998, Nature, 393, 763.
[23] Ellis, J., et al. 2000, ApJ, 535, 139.
[24] Albert, J., et al. 2008, Phys. Lett. B, 668, 253.
[25] Aharonian, F., et al. 2008, Phys. Rev. Lett., 101, 402.
[26] Abdo, A.A., at al. 2009, Science Online, February 19.
[27] Fazio, G.G., & Stecker, F.W. 1970, Nature, 226, 135.
[28] Coppi, P., & Aharonian, F. 1997, ApJ, 487, L9.
[29] Aharonian, F., et al. (H.E.S.S. Collaboration) 2006, Nature, 440, 1018.
[30] Albert, J., et al. (MAGIC Collaboration) 2008, Science, 320, 1752.
[31] Stecker, F.W., Baring, M.G., & Summerlin, E.J. 2007, ApJ, 667, L29;
[32] Aharonian, F., Khangulyan, D., & Costamante, L. 2008, arXiv:0801.3198.
[33] De Angelis, A., Roncadelli, M., & Mansutti, O. 2007, Phys. Rev. D, 76, 121301.
[34] Abraham, J., et al. (Pierre Auger Collaboration) 2007, Science, 318, 939.
[35] Masso, E., & Toldra, R. 1995, Phys. Rev. D, 52, 1755;
[36] Chang, S., Tazawa, S., & Yamaguchi, M. 2000, Phys. Rev. D, 61, 084005
[37] Turok, N. 1996, Phys. Rev. Lett. 76, 1015; Svrcek, P., & Witten, E. 2006, JHEP, 0606, 051.
[38] Carroll, S.M. 1998, Phys. Rev. Lett., 81, 3067.
[39] Kronberg, P.P. 1994, Rept. Prog. Phys., 57, 325;
[40] De Angelis, A., Persic, M., & Roncadelli, M. 2008, Mod. Phys. Lett. A, 23, 315.
[41] Hooper, D., & Serpico, P.D. 2007, Phys. Rev. Lett., 99, 231102;
[42] Kneiske, T.M., Bretz, T., Mannheim, K., & Hartmann, D.H. 2004, Astron. Astrophys., 413, 807.
[43] Kamionkowski, M. 1995, in *The Gamma-Ray Sky with CGRO and SIGMA*, ed. M. Signore, P. Salati, & G.Vedrenne (Dordrecht: Kluwer), 113.
[44] Willis, T. 1996, Ph.D. thesis, Stanford Univ.
[45] Kribs, G.D., & Rothstein, I.Z. 1997, Phys. Rev. D, 55, 4435.
[46] Schwartz, R.N., & Townes, C.H. 1961, Nature, 190, 205.
[47] Howard, A.W., et al. 2004, ApJ, 613, 1270.
[48] Eichler, D., & Beskin, G. 2001, Astrobiology, 1, 489.
[49] Holder, J., et al. 2005, Proc. 29th ICRC, Pune.
[50] Armada, A., et al. 2005, in *Neutrinos and Explosive Events in the Universe*, Proc. 14th Intl. School of
[51] Kifune T. 1999, ApJ, 518, L21.
[52] Amelino-Camelia, G. 2008, arXiv:0806.0339.
[53] Heitler, W. 1960, *The Quantum Theory of Radiation* (Oxford: Oxford University Press).

# 5 Technologies for Cherenkov telescopes.

The scientific requirements discussed in previous paragraphs address the experimental TeV astronomy towards three well specific directions:
- To improve the sensitivity of one order of magnitude with respect to the present Cherenkov telescopes in the energy band from 200 GeV to 10 TeV by covering a large physical area with several tens of Cherenkov telescopes;
- To lower the energy threshold up to few tens of GeV, connecting the Cherenkov spectra with the spectra measured by the space-based instruments (e.g. Fermi, NASA Space mission);
- To extend the energy band up to 100 TeV and more exploring the astrophysical acceleration processes and search for unexplored and unpredicted phenomena.

The following figure gives a visual sketch of this so large energy interval (four order of magnitude at least):

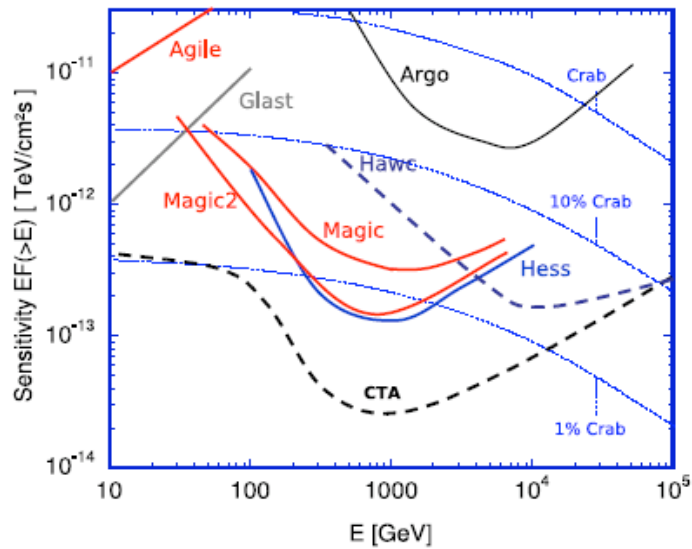

*Fig. 5.1:* *Sensitivities of some present and future HE gamma detector, measured as the minimum intensity source detectable at 5 sigma [1]. The performance for EAS and satellite detector is based on one year of data taking; for Cherenkov telescopes it is based on 50 hours of data.*

Such ambitious program requires a big effort in using and improving the technologies more suitable for reaching the objectives in a reasonable time interval (< 10 years) and with reasonable costs (~150 Meuro). The recent decision of ESRI [2] of including the European Cherenkov Telescope Array (CTA) [3] in the first 8 projects in "Physics Sciences and Engineering" for the roadmap 2008 gives realistic perspectives of a next start of the design study of this project.

## *5.1 Which telescopes in the array?*

In the atmospheric showers originated by a primary Gamma ray, the Cherenkov light intensity is almost proportional to the Gamma ray energy. In general, large pupil mirrors are needed to trigger low energy Gamma-rays, while small mirrors are sufficient enough to trigger high energy Gamma-rays. Moreover, due to the very low values of the Gamma Ray fluxes at high energy, future

Cherenkov telescopes must be able to catch events with a core (impact point) very far from the telescope position, reaching, in this way, effective areas of the order of millions of square meters; to trigger far showers, imaged at large off-axis angles, Cherenkov telescopes must be provided by sufficiently large fields of view.

Optical dish diameters and field of view are the first parameters to be considered before analyzing other specific aspects as optical design, mirrors structure, focal camera sensors and electronics.

Due to the widespread of the spectral band, and to the phenomenological constraints three different specific set of telescope with different characterists are envisaged for future Cherenkov telescopes arrays to cover the energy interval from 10 GeV up to 100 TeV and more:

- very low energies below ( < 200 GeV) need very large optical dishes ( >20 m diameter ), but a small field of view (~ 2° full) is sufficient to image showers impacting till distance less than 120 m; beyond this distance also twenty meter pupils and more are not enough large to trigger the faint Cherenkov light produced by Gamma Ray of several tens of GeV;
- in the extreme energy range 10-100 TeV, the intensity of the Cherenkov signal allows to image very far showers with small (< 10 m) optical dishes provided by a sufficient wide angle (~8° full);
- intermediate energy band (0.2–10) TeV will be covered by telescopes with average characteristics for what concerns dish diameter and field of view.

## 5.2  *Optical design.*

The present telescope used by MAGIC [4], HESS [5] and VERITAS [6] are based on the Davies-Cotton (DC), single mirror design, this choice has allowed the realization of telescopes with good performance and moderate costs mainly for the mirror system. An alternative possibility is the Schwarzschild-Couder (SC) telescope design which uses primary and secondary mirrors making possible a better angular resolution, a larger field of view and a decrease of the focal length. In terms of costs the SC is more expensive for what concerns mirrors but allows smaller cameras with pixels size compatible with MAPMT (Multi Anode Photomultipliers), for this reason the total cost of SC system could be at the end competitive with the DC model.

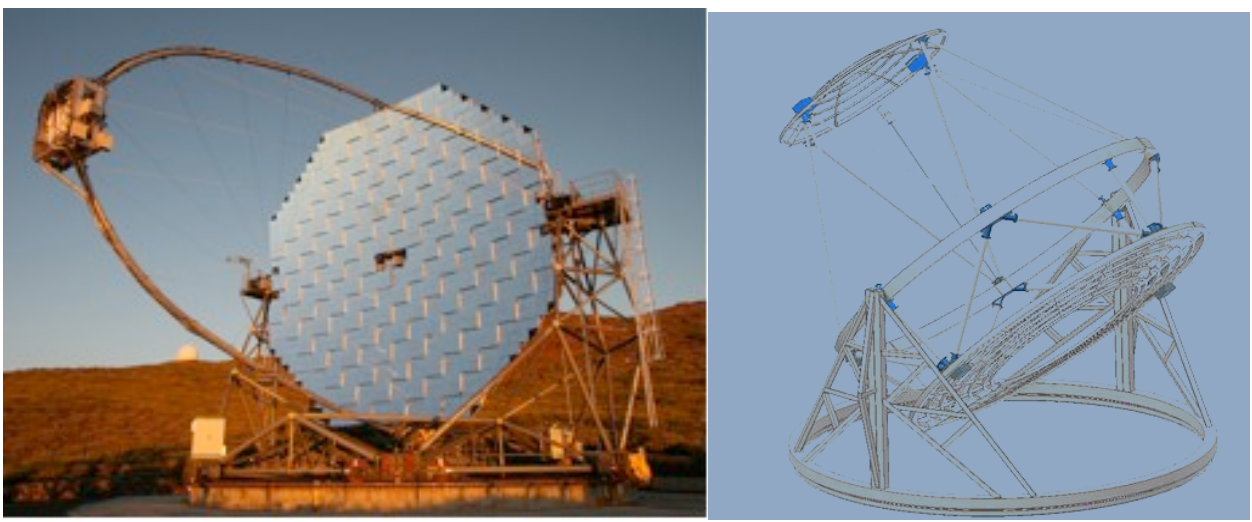

*Fig. 5.2.1: Davis Cotton Telescope, MAGIC [4], left panel. Schwarzschild-Couder SC Serrurier Design , AGIS, right panel [7].*

## 5.3 Mirror technology.

Since it is foreseen that for CTA it will be necessary to produce about 10000 m$^2$ of mirrors it is important to use a fast technique of production. Different techniques of production are able to permit the manufacturing of long radius of curvature segments (to be used for the assembly of the primary mirrors) and relatively short radius of curvature segments (necessaries for example for the manufacturing of the secondary mirrors of large field Cherenkov telescopes).

### 5.3.1 Panels made by cold slumping of thin glass segments.

If the radius of curvature requested for the panels is long (i.e. 20-30 m) and hence the sag of the panel surface is small, it is possible to take advantage from the flexibility of very thin glass sheets (1-2 mm) and create a very stiff and lightweight panel. In this case the shape is imposed using a high quality convex optical master. By means of a vacuum suction applied between the master and the thin glass sheet, the glass itself is fully forced against the master surface. At this point it is applied on the curved glass a epoxy glue and then above it a glass foam or a Silicon Carbide foam having a thickness of some centimetres. The side in contact with the glass of this foam will be preformed in a concave way so to conform to the bended glass sheet. Small differences, in shape between glass and foam, will be compensated by the glue. To complete the manufacturing process another flat glass sheet will be glued in the same way on the upper surface of the foam so to form a stiff and sandwiched panel. After the curing of the glue the vacuum suction will be removed and the panel will be ready for the aluminization of the glass concave surface. With this approach, that is an improved version of the technique already used for the production of the MAGIC II panels [8, 9], it will be possible to reach a very fast production of the panels, also characterized by very low weight.

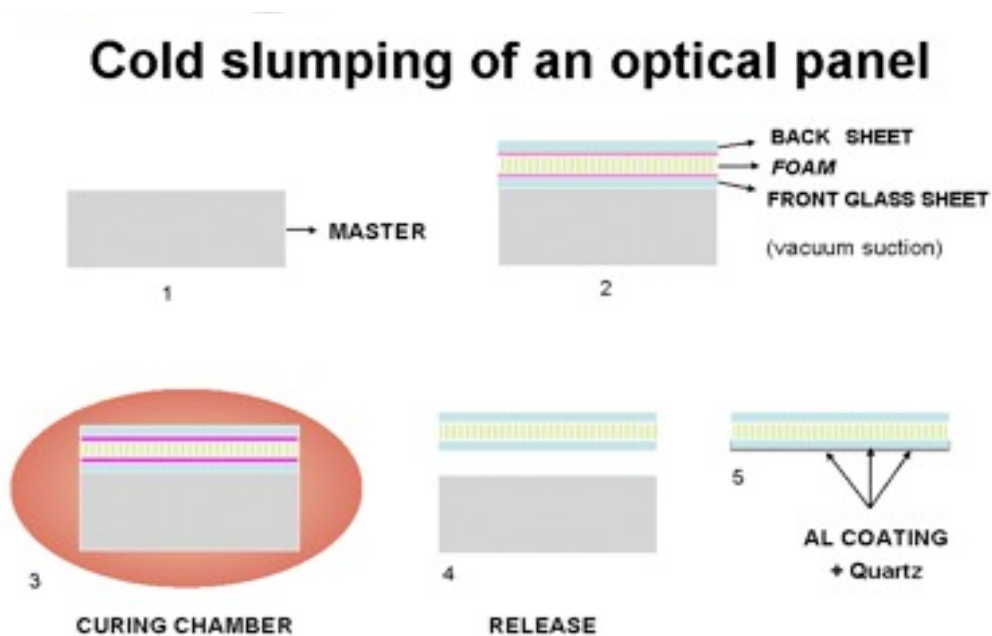

*Fig. 5.3.1: Cold slumping technique for the production of optical panels.*

*5.3.2 Panels made by hot slumping of glass segments.*

If the radius of curvature of the panels is shorter (10-15 m) it is no longer possible to bend adequately the glass sheets. It is hence necessary to add a step to the previously described technique. This version of the technology [10] is based on a convex optical mould made in ceramic and able to withstand high temperatures. To change the shape of an initially flat sheet of glass having thickness of few mm the glass itself is placed onto the mould and then a suitable thermal cycle is applied to reach a temperature in which the glass is softened and its shape is changed copying the master shape. After the thermal cycle the vacuum suction is again used to force the glass sheet against the mould. It must be noticed that after the thermal cycle the glass sheet will have a shape very near to that of the mould and hence the vacuum step is used only to remove small shape differences between the two surfaces. This means also that there will be essentially no spring-back effect of the glass since its shape is thermally imposed. Following the procedure, after that the vacuum suction is applied on the glass, the epoxy glue is applied and then the preformed foamed material is placed in contact with it. Again, on the upper part of it will be glued another flat glass panel to create a sandwich structure. The panels produced will be quite rigid and lightweight, making possible to reduce the requirements on the mechanical structure of the telescope and lowering the inertia of the system.

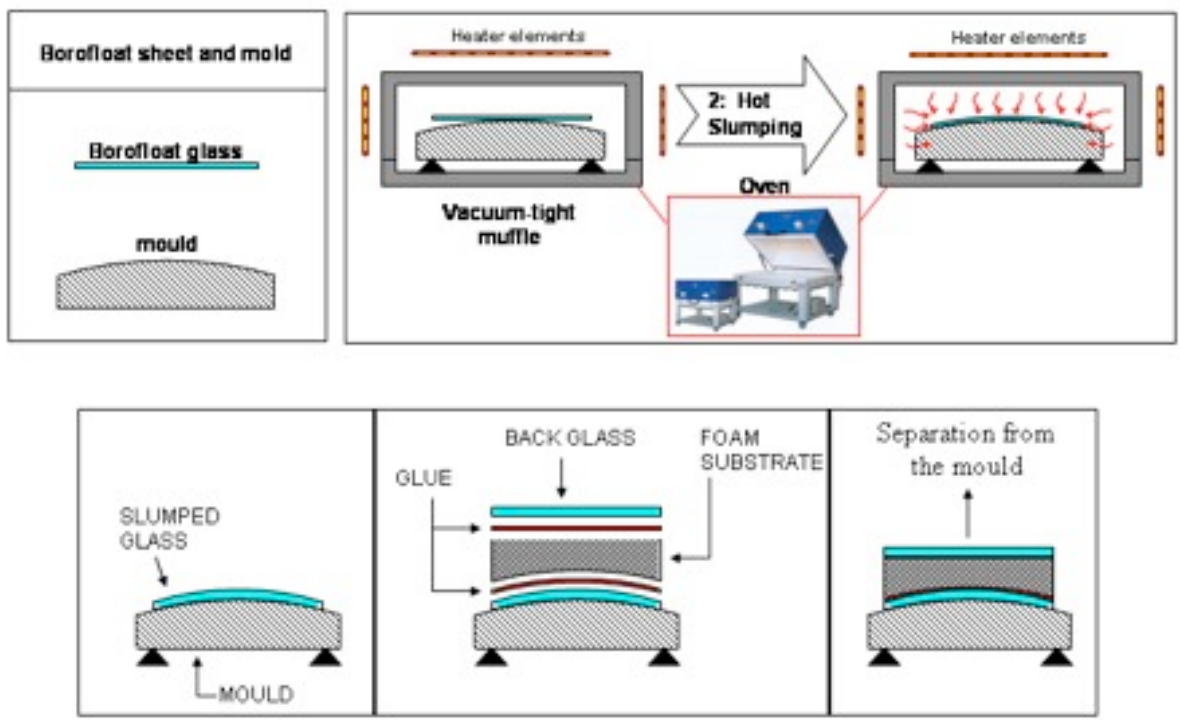

***Fig. 5.3.2:*** *Hot slumping technique for the production of optical panels.*

## *5.4 Sensors for the focal camera.*

A focal camera suitable for Cherenkov telescopes should have the following characteristics:

- High segmentation in pixels of few arc-minutes each-one,
- Very fast response, the nanosecond or better is the time resolution necessary to keep negligible the night sky background and to obtain first order indication on the internal time structure of the Cherenkov flash.
- A good efficiency and a low level of intrinsic noise, to gain in the lower energy threshold.
- Curved focal surfaces, to fit the ideal focal surface mainly in the cases of large field of view.

### *5.4.1 Photo Multiplier Tubes.*

The **Photo Multiplier Tube** (PMT) is the sensor more suitable for matching the Cherenkov camera requirements, in the present Cherenkov telescopes the typical focal camera is constituted by a set of single PMTs packed together and each-one constituting a single pixel.

Limits in the use of one pixel PMTs is in the quantum efficiency of traditional photo-cathodes (<~25% at 400 nm), in the loss of continuity of the focal surfaces, in the pixel field of view achievable, in general too large, and in the costs which are of the order of several hundreds of Euros per pixel.

To gain in quantum efficiency, **Silicon PMs** have been studied in the recent past, but the technology is not mature for a present use. Nowadays, they are not yet completely responding to Cherenkov camera requirements, mainly for two unresolved problems:

- pixels dimensions are less than 100 μm, while the order of a few millimeters is the size more convenient for matching the typical focal lengths of Cherenkov camera obtaining sky pixels of a few of arc minutes;
- the intrinsic noise is too high (orders of magnitude more) with respect to the PMT noise.

The **Multi Anode PMTs** (MAPMTs) with UV transmitting windows and Super or Ultra bialkali photo cathodes are today on the shelf and they represent the sensors more convenient for Cherenkov Camera. MAPMTs are characterized by:

- peak quantum efficiency > 40% at 350 nm in the case of the ultra bialkali photo cathode [11];
- number of pixels 4x4, 8x8 or 16x16 with pixel size of mm order, allowing pixel field of a few arc minutes;
- square dimensions of 30-50 mm side with a small dead border area, the MAPMT size allows a curvature correction with steps of the order of few tens arc minutes for 10 m focal length;
- gain up to $10^6$, allowing the use with various electronics techniques.

Furthermore, the cost/pixel of the MAPMT solution is certainly competitive compared to the single pixel PMT.

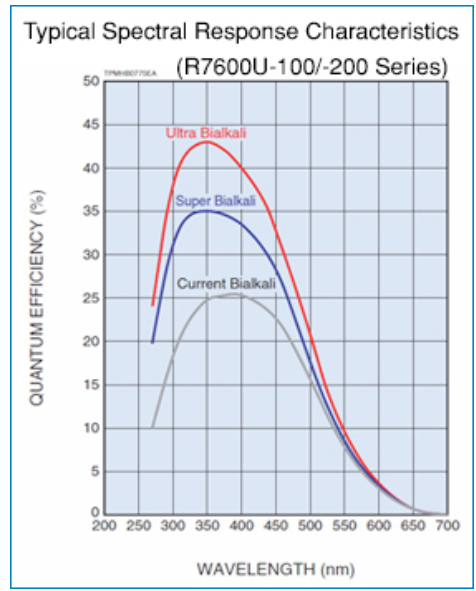

*Fig. 5.4.1:* Spectral Response for Super and Ultra Bialkali photo cathodes in MAPT Hamamatsu R7600 [11].

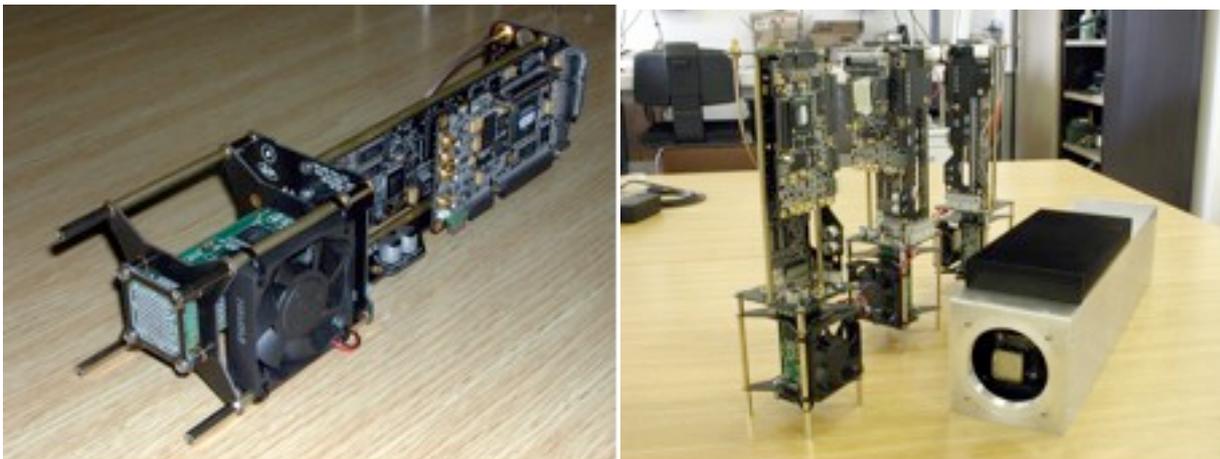

*Fig. 5.4.2:* (left panel) is shown an apparatus, realized in IASF-Palermo, to test and qualify MAPT Hamamatsu R7600, GAW experiment [12]. Right panel shows the apparatus within its mechanical box.

### *5.4.2 Solid State Detectors.*

Another kind of detectors suitable for the focal plane of Cherenkov telescopes can come from the silicon industry, in fact they have been recently developed two solid state detectors named:

- Single Photon Avalanche Diode (SPAD),
- Silicon Photon Multiplier (SiPM)

both operating in photon counting regime in continuous mode, and characterized by both a good quantum efficiency in the visible and ultra fast response (tens to hundreds of picoseconds).

The **Single Photon Avalanche Diode (SPAD)** [13, 14] is a silicon sensor able to detect single photon events. It is essentially an avalanche photodiode that, biased above breakdown, remains quiescent until a carrier, generated either thermally or by a photon, triggers an avalanche. A quenching circuit (active or passive) extinguishes the avalanche and makes the pixel ready to detect another photon [15, 16].

The **Silicon Photon Multiplier SiPM** [17], as well as the SPAD, is a photodetector operated in Geiger mode, and it is constituted by hundreds/thousands of pixels, and the discharge is quenched by a small polysilicon resistor (passive quenching) in each pixel. The micro-pixels, having typical sizes of tens of microns, are built on a common substrate (Fig.5.4.3).

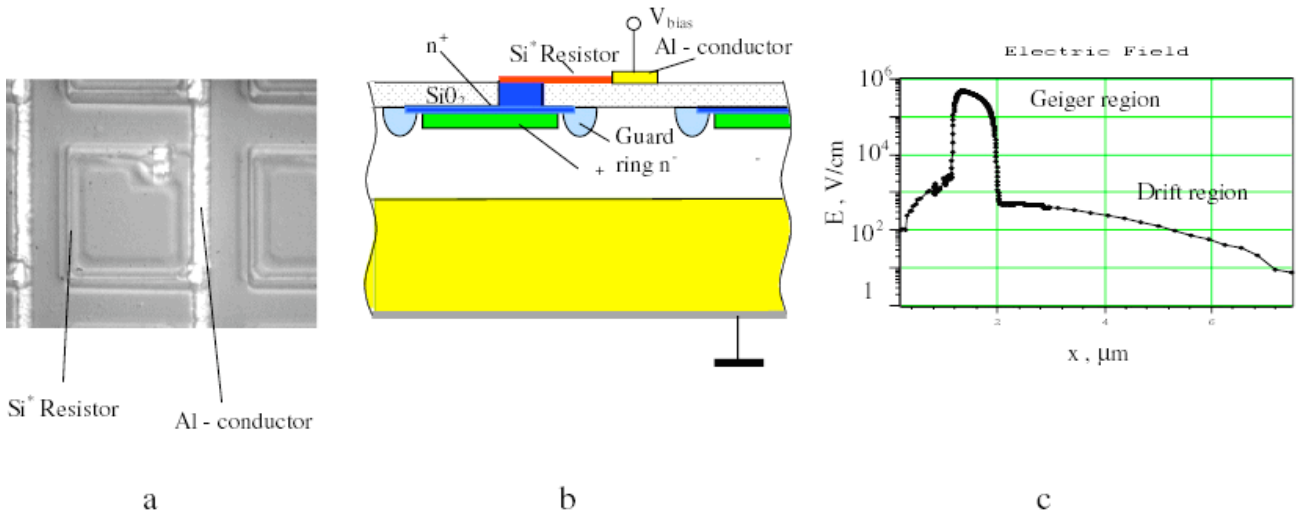

*Fig. 5.4.3:* (a) SEM microphotography of a SiPM micro-pixel, (b) schematic section of the micro-pixel (c) electric field dependence through a vertical section of the micro-pixel.

The independently operating pixels are connected to the same readout line; therefore the combined output signal corresponds to the sum of all fired pixels. It reaches an intrinsic gain for single photoelectron of $10^6$, comparable to that of vacuum phototubes (PMTs).

The SiPM device, besides the same SPAD time tagging characteristics, has the advantage of a larger area than that of the SPAD and can detect more photons simultaneously.

The performances of SiPMs are very interesting, in particular if compared to PMT tubes in fact they can show same gains at low operating voltages (~ 30V), operational stability, insensibility to the magnetic field, single photon detection, possibility to operate at room temperature although best noise performances are obtained at lower temperatures (up to -30°C). The SiPM dynamic range is directly proportional to the number of micro-pixel in the array; for this reason it is important to build very small but highly efficient, micro-pixels.

Currently, the limiting factor for the SiPM in the single photon detection mode is the noise rate due to the dark current, typically few MHz/mm$^2$ (room temperature) or less than 1 KHz/mm$^2$ (~ -40°C). Due to the random properties of the noise signals, increasing the threshold of the output signal to value corresponding to two (or more) simultaneous pulses reduces the noise rates by order(s) of magnitude. At the moment SiPMs are manufactured in a format that ranges from 1 x 1 mm$^2$ to 5 x 5 mm$^2$ with thousand micro-pixels. To cover a larger area as that of PMT tubes some manufacturers have developed mosaics of SiPMs. In particular the SensL and the ST Microelectronics have already realized arrays of 8 x 8 SiPMs being each SiPM 4 mm$^2$ reaching a sensitive area of 32 x 32 mm$^2$ and obtaining a so called multi-anode architecture. In Fig. 5.4.4 is shown a 16 elements SiPMArray manufactured by SensL.

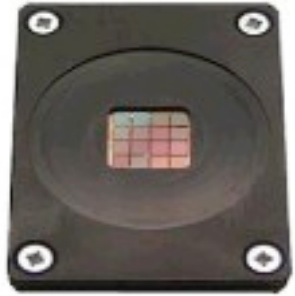

*Fig. 5.4.4:* SiPMArray constituted by 4 x 4 SiPM of 3x3 mm$^2$ (Position Sensitive Multi Anode High Gain APDS).

Some new STM SiPM and Hamamatsu Multi Pixel Photon Counter (MPPC) devices (Fig. 5.4.5) have been already characterized at the INAF–Catania Astrophysical Observatory.

The STM SiPM 100-cells has dimensions of 0.5x0.5 mm$^2$ with each cell squared and with a 50μm/30μm side over active area ratio giving a 36% of fill factor, while the Hamamatsu 100-cells has a pitch of 100 μm over a squared millimeter giving a fill factor of 78.5 %.
Both types of sensor are biased slightly above the breakdown by an overvoltage (around 10% for the STM and around few percent for the Hamamatsu).

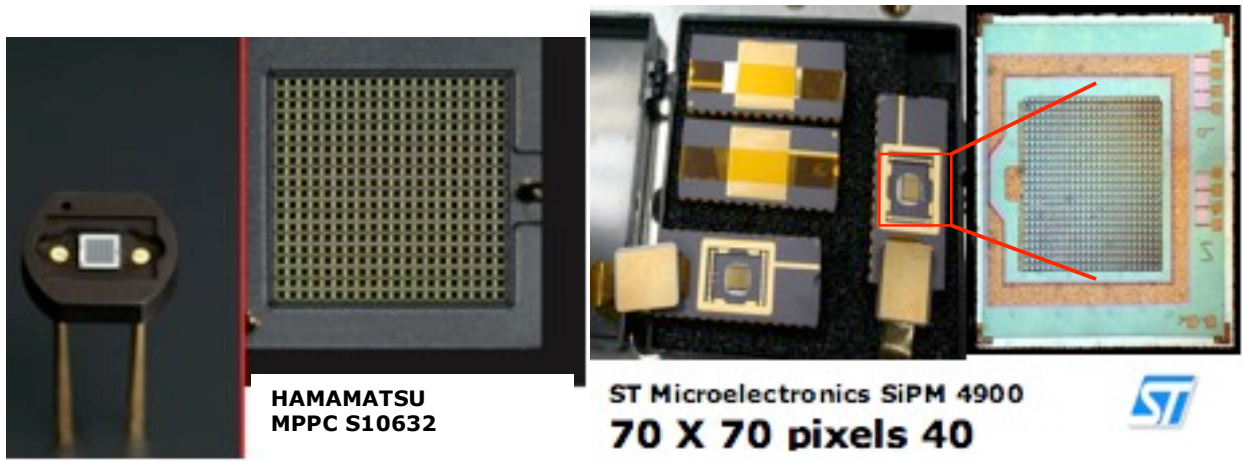

*Fig. 5.4.5: (Left panel)* 100-cells Hamamatsu Multi Pixel Photon Counter (MPPC) device. *(Right panel)* 100-cells STM SiPM device. Both sensors have been operated using a bias over-voltage of about 10% for the STM (breakdown voltage is about 29.5 Volts) and of about few percent for the Hamamatsu (breakdown voltage is about 68,6 V).

The measured PDE using the so called "Counting method" [17], consisting in counting each produced event and accounting for afterpulses and other spurious pulses, for both devices is shown in fig. 5.4.6. As can be seen a PDE peak of about 30% is obtained at the wavelength of about 420 nm (Cherenkov light).

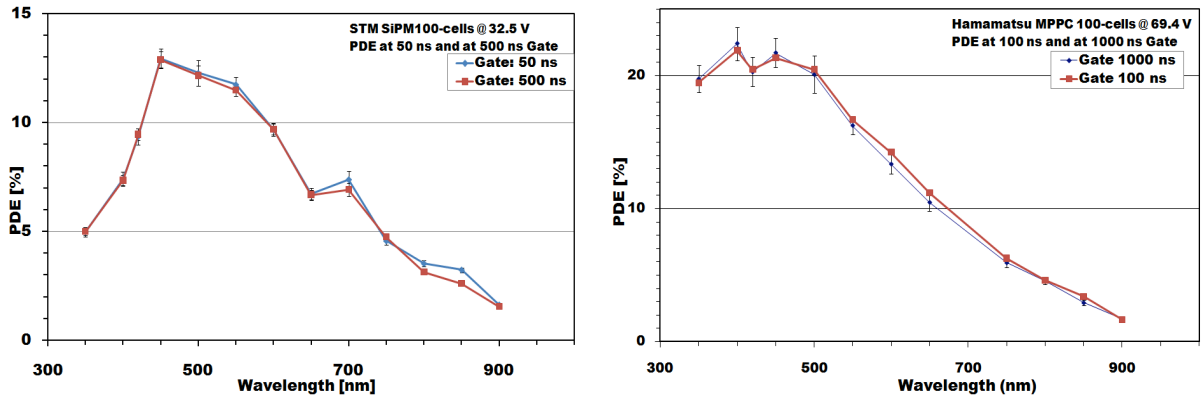

*Fig. 5.4.6: (Left panel) The PDE of the STM device biased at 32.5 V. (Right panel) The PDE measured for the Hamamatsu 100 cells biased at 69.4 V.*

We have to stress that this PDE values are about half of the typical SPAD PDE, essentially due to the small fill factor of a SiPM. But this value can be improved by using some treatment of the single element sensitive area.

Future improvements are necessary: larger pixels to have grater fill factor and better structure or some sort of cooling to lower the dark current. We are confident that SiPM devices in the near future can be considered valid substitution of PMT devices, especially for applications that require a multi-anode configuration.

## 5.5 Electronics.

The pixel front-end electronics successfully used for the H.E.S.S.-1 and -2 experiments and in MAGIC-I and MAGIC-II experiments is based on the conversion of the electrical pulse (integrated charge) to the so-called ADC-counts. The relative measurement provided by ADC counts is then converted to equivalent pe (photo-electrons) knowing the mean charge produced by a single pe. The mean pe charge can be statistically measured with the SER (single electron response) technique. Both the experiments have improved, in the last camera-telescopes versions, the performance of the analogue sampling with new FADCs (Fast Analog to Digital Converter), based on ASIC (Application Specific Integrated Circuit) switching-capacitor chips exhibiting a better timing (up to 4Gsps) and amplitude resolution (up to 12 bits) and contemporary reducing the power consumption as low as 300 mW/chip. Further developments have already started to make future chips even more flexible and reliable.

An alternative technique to the analogical technique, never used up to now by any experiment, is the Single Photon Counting technique (SPC) [18]. In this case, the pixel front-end electronic is basically constituted by a fast preamplifier, a discriminator, and a pulse former. In order to reproduce faithfully the anode-pixel pulses, a high-speed linear preamplifier is directly connected to each anode. An adjustable threshold discriminator shapes the amplified pulse to a standard digital logic level. The resulting digital signal sampled at a given frequency represents the pes-counts. This detection method has been studied and will be used in the GAW project, a pathfinder experiment for a new generation of high sensitivity large field Imaging Atmospheric Cherenkov Telescope.

Merits and limits obviously exist for both the methods cited. Among many of them, the main advantages of Single Photon Counting, that specularly represent the limits of the Charge Integration method and vice versa, can be summarized as follow:

- minimizing changes in the count rate with respect to variations in the supply voltage, without sacrificing the signal-to-noise ratio. This means that the SPC mode ensures high stability even when the gain of the photo-sensor varies (for example: ambient temperature), as the gain is a function of the supply voltage;

- allowing the detection of very faint light (as demonstrated recently [18]). This allows the trigger of events characterized by a lower number of photo-electrons. So, it will contribute to the general performance of the telescope in the entire spectrum by lowering the energy threshold and, at high energy, by triggering very far showers, with the positive consequence of enlarging the telescope effective area;

- realizing simple front-end electronics.

whereas the limits are:

- signal pile-up for intense light (limits in dynamic range)

- the front-end electronics must be integrated in a customized ASIC

## *5.6 Data Handling, Data Processing, Data Access.*

Cherenkov telescopes are more complicated than other astronomical instruments that take images of the sky directly. They require an extensive data processing in order to obtain from the Cherenkov images the parameters of the primary gamma ray. Depending on the emphasis in the data analysis there is a wide range of selection parameters, all resulting in different effective detection areas and instrument characteristics. Effective detection areas also depend on the zenith angle, orientation relative to the Earth magnetic field, etc. Effective areas determination, hadron-photon separation and background subtraction are very critical for these instruments. Scientific products are typically obtained after a complicated and not-standard reduction process. Such a complicated data processing and analysis can be carried on in an *experiment* context such as MAGIC, HESS, VERITAS and CANGAROO (see par. 1.4). However, CTA will be operated as an observatory and the community of users of the CTA data will be not limited to those most familiar with the instrumentation, but will involve also a larger community of astroparticle physicists and astronomers from different fields. This will imply that both the CTA data and the data analysis tools should be distributed in a format that should be readily accessed by external observers. Moreover, CTA data will be complementary to those obtained with other astronomical facilities at different band so, an easy access and comparison of all these data will be mandatory in order to maximize the scientific output. Minimum requirements to allow an easy and effective access to CTA data are:

- The CTA data should be processed adopting standard procedures, standard software and calibration.

- The archival CTA data and the results of the standard data processing should be available in a format that will be easily transportable and compatible with the standard data formats used in astronomy (e.g. FITS) and compliant with the International Virtual Observatory (IVOA) requirements.

- Data analysis software tools should be developed and maintained in a fashion easily accessible to a larger community.

- The CTA Data Archive should be designed in order to allow an easy access in a homogeneous way to CTA data in order to compare them with a variety of sky maps, catalogues and scientific products at different wavelengths and from different experiments.

The large expertise in data handling and archiving present in many INAF institutes guarantees a major contribution to CTA to all these topics by providing requirements, defining specification, developing prototypes. In particular, INAF contribution will be very important in the field of multiwavelength and multi-messenger analysis.

INAF groups have participated to the design and development of data handling and archiving systems for several Space Missions (e.g. BeppoSAX, XMM, SWIFT, AGILE, Planck) as well as for ground-based instruments for astronomy (optical telescopes (e.g. REM, TNG, LBT) and radio telescopes) and cosmic rays (e.g. EAS-TOP, ARGO-YBJ, ULTRA, GAW). The management of databases for astrophysical observatories has been also a duty fulfilled at INAF since many years. INAF is also participating to the ASI Science Data Centre[10] scientific and technical activities. Through the collaboration between INAF and ASDC, it will be possible to access a large number of high energy missions databases such as AGILE, Fermi and Swift, as well as past experiments like Beppo-SAX, in order to cross-calibrate the CTA and to perform multi-wavelength studies.

## *5.7 Observatory Operations.*

CTA will be operated, for the first time in this field, not as an experiment but as a user facility with a public Guest Observer Program. This *new* (for this energy band) approach raises issues such as operation of the facility as well as data acquisition and dissemination. The large flexibility provided by the CTA array also raises new challenges concerning the management of the observatory and the scheduling of observations. In the design study phase, a proper model for the technical and scientific operation of the instruments has been specifically developed in order to have the best use of the facility and to operate it like an open observatory. Technical and scientific operations for CTA mean to operate a large number (>100) of telescopes possibly in two different sites in order to:

- Transform proposals in scientific observations by operating the observatory in a robotic or service mode;
- Manage different input (e.g. ToO) through a dedicated scheduler to optimize science;
- Operate the array(s) or part of it;
- Acquire and manage all the information necessary to obtain scientific data (e.g. atmospheric data);
- Distribute data to Guest Observers
- Archive data to allow easiest data mining

Many INAF groups (e.g. OA Roma, OA Bologna, OA Brera) have a very good expertise in the management of both space- and ground-based facilities. The INAF contribution to the CTA for this task includes the development of system for proposal treatment, observation schedules management, array operation, data handling as well as the treatment of final data products in a Data Center.

In INAF indeed, there is also a deep expertise in both remote and automatic control of operations of an astronomical observatory (e.g REM). This expertise can be applied to CTA in case the project will evolve toward robotic operations.

---

[10] http://www.asdc.asi.it/

# References

[1] De Angelis, A., et al., 2008, Nuovo Cimento, Vol. 31, N. 4, pp. 187-245.
[2] ftp://ftp.cordis.europa.eu/pub/esfri/docs/esfri_roadmap_update_2008.pdf
[3] http://www.cta-observatory.org/
[4] http://wwwmagic.mppmu.mpg.de/
[5] http://www.mpi-hd.mpg.de/hfm/HESS
[6] http://veritas.sao.arizona.edu
[7] http://gamma1.astro.ucla.edu/agis/images/6/6c/AGIS_mechanics_v5.pdf
[8] D. Vernani; R. Banham; O. Citterio; F. Sanvito; G. Valsecchi; G. Pareschi; M. Ghigo; E. Giro; M. Doro; M. Mariotti, Proc SPIE 7018, 2008
[9] G. Pareschi; E. Giro; R. Banham; S. Basso; D. Bastieri; R. Canestrari; G. Ceppatelli; O. Citterio; M. Doro; M. Ghigo; F. Marioni; M. Mariotti; M. Salvati; F. Sanvito; D. Vernani, Proc SPIE 7018, 2008
[10] R. Canestrari; M. Ghigo; G. Pareschi; S. Basso; L. Proserpio, "Investigation of a novel slumping technique for the manufacturing of stiff and lightweight optical mirrors", Proc SPIE 7018, 2008
[11] http://jp.hamamatsu.com/resources/products/etd/eng/html/pmt_003.html
[12] http://gaw.iasf-palermo.inaf.it
[13] F.Zappa et al., , Sens. Actuators A, vol. 140, pp. 103-112, **2007**.
[14] Cova, S.; Ghioni, M.; Lotito, A.; Rech, I.; et al.. Journal of Modern Optics **2004**, 51 (9), 1267-1288
[15] Cova S.; Ghioni M.; Lacaita A.; Samori C.; Zappa F., Applied Optics, **1996**, Vol. 35, No. 12.
[16] Golovin V.; Saveliev V., Nucl. Instrum. and Meth. A, **2004**, 518, 560-564.
[17] P. Finocchiaro et al, IEEE T. E. D., **55**, No. **10**, p. 2757 (2008)
[18] Catalano, O., Maccarone, M.C., Sacco, B., 2008, Astroparticle Physics, Volume 29, Issue 2, p. 104-116